\documentclass[11pt]{article}
\usepackage[a4paper,margin=1in]{geometry}
\usepackage{authblk}  % author affiliation 패키지
\usepackage{setspace} % optional (줄간격 조절)
\usepackage{graphicx,amsmath,amssymb}
\usepackage{hyperref}
\usepackage{xcolor, soul}
\hypersetup{
    colorlinks=true,
    citecolor=red
}  
\title{Inhomogeneous mixing: From microscopic dynamics to mesoscopic staircases}

\author[1,*]{T. Long}
\author[2,*]{M. J. Choi}
\author[3,*]{P. H. Diamond}

\affil[1]{Southwestern Institute of Physics, Chengdu China}
\affil[2]{Korea Institute of Fusion Energy, Daejeon, Republic of Korea}
\affil[3]{Departments of Astronomy \& Astrophysics and Physics, University of California, San Diego, CA, United States of America}

\affil[*]{Corresponding authors: longt@swip.ac.cn, mjchoi@kfe.re.kr, pdiamond@ucsd.edu}

\date{\today} % or leave empty with {}

\begin{document}
\maketitle

%%%% Abstract text to be placed here %%%%%%%%%%%%
\begin{abstract}
Inhomogeneous mixing and the consequent mesoscopic layered structure have been observed in many physical systems, including magnetically confined fusion plasmas. 
Especially, in plasmas, mixing can be enhanced through turbulence spreading by intermittent coherent structures (blobs/voids), or suppressed due to the formation of transport barriers (sheared zonal flows). 
Interestingly, blobs/voids and zonal flows are not independent, and they can co-exist in a state of inhomogeneous mixing, often called the $\mathbf{E} \times \mathbf{B}$ staircase.
In this paper, we first introduce recent experimental progress on the physics of blobs/voids: how turbulence spreading by blobs/voids occurs, the consequences of enhanced turbulence spreading for the scrape-off layer (SOL) power decay length, and the interaction between blobs/voids and zonal flows.
Then, we provide a brief review of experimental results on staircases, or more generally layered mesoscopic transport barriers. 
Staircases are often elusive in experiments, requiring integrating multi-diagnostic data, utilizing high-dimensional diagnostics, or extracting hidden information from signals for their identification. 
Our understanding is still incomplete.  
This paper serves as an initial step toward applying insights gained from inhomogeneous mixing due to blobs/voids to the understanding of a staircase. 

%Such a structure of inhomogeneous mixing is often called the $\mathbf{E} \times \mathbf{B}$ staircase, or more generally mesoscopic transport barriers. 
%In this paper, recent experimental progress on revealing complex roles of blobs/voids would be introduced, revealing various roles of coherent structures and the identification and characterization of mesoscopic staircase   
%Although there are a number of experiments which report the observation of a staircase, their physics is still 
%In this paper, we first introduce recent experimental progress in the physics of blobs/voids: how turbulence spreading by blobs/voids occurs, the consequences of enhanced turbulence spreading on the power decay length, and the interaction between blobs/voids and zonal flows. 
%Then, 

\end{abstract}
%%%%%%%%%%%%%%%%%%%%%%%%%%%

%%%%%%%%%%% Insert the texts which can accomdate on firstpage in the tag "fmtext" %%%%%
%\maketitle
%%\begin{fmtext}
%%\end{fmtext}
%
%%%%%%%%%%%%%%%% End of first page %%%%%%%%%%%%%%%%%%%%%

\section{Introduction}

\begin{figure}[!h]
%\centering\includegraphics[width=5.0in]{PD01_rev.png}
\centering\includegraphics[width=6.0in]{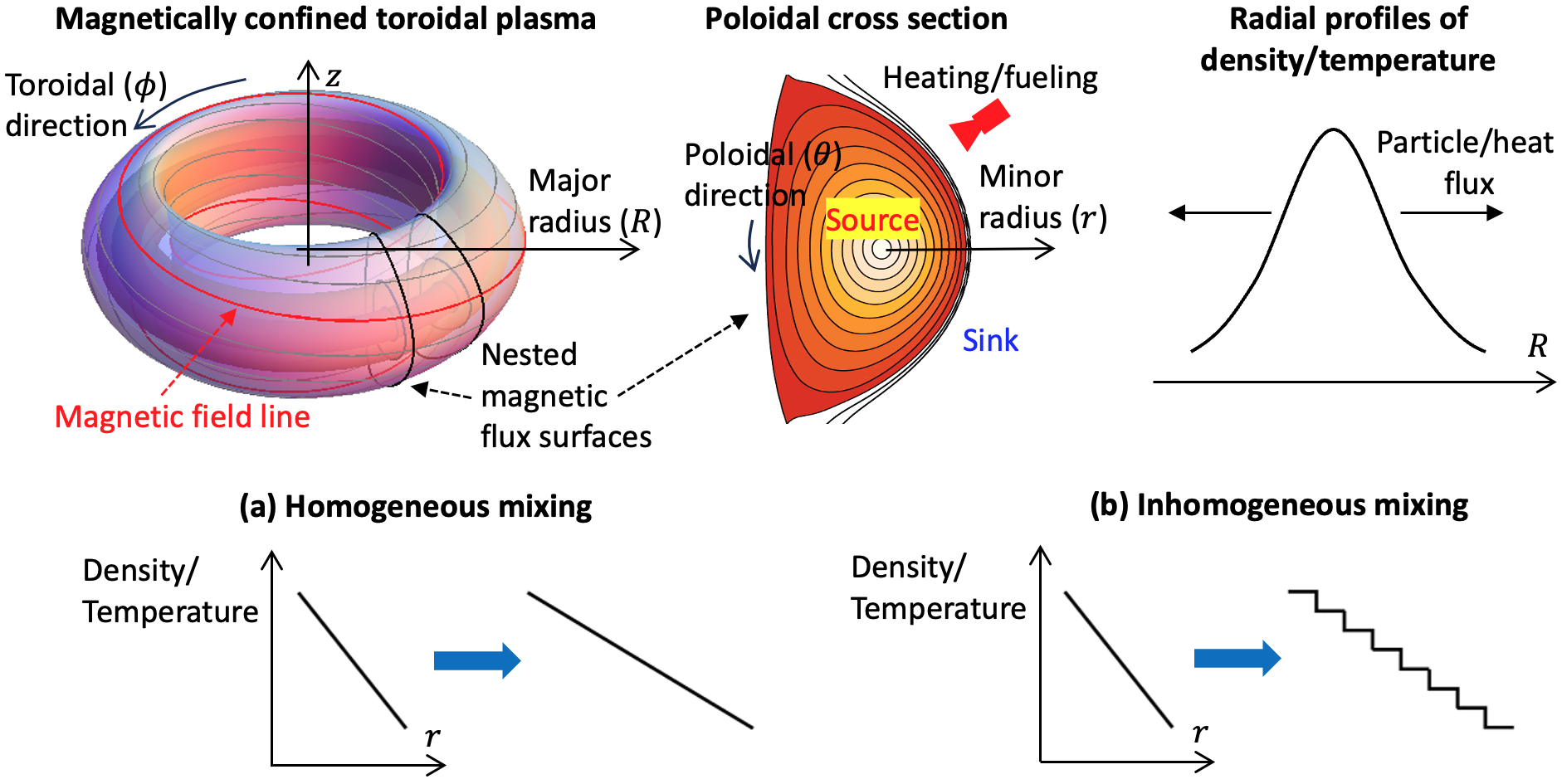}
\caption{Toroidal plasma confined by toroidal and poloidal magnetic fields. Plasma is heated and fueled by external devices, forming the peaked density/temperature profile for fusion reactions. Plasma instabilities, growing due to the gradient of profiles, can develop into plasma turbulence which mixes density/temperature. Schematic diagram of (a) homogeneous mixing and (b) inhomogeneous mixing of density/temperature. Panel (b) shows a staircase in plasma density/temperature, with flat regions representing well-mixed regions and jumps representing transport barriers to mixing.}
\label{fig:PD01}
\end{figure}

Inhomogeneous mixing is a transport or relaxation process which is not uniform. 
So, in Figure~\ref{fig:PD01}, case (a) depicts homogeneous mixing, while case (b) shows inhomogeneous mixing.
Note there, the mixing zones don’t overlap. 
As suggested by Figure~\ref{fig:PD01}, inhomogeneous mixing of a profile will quite naturally produce layered or staircase structures on mesoscales or macroscales. 
Such layered structure is the focus of this collection, and staircases in magnetized plasma are the foci of this paper. 
Inhomogeneous mixing and staircases form in many physical systems, including saltwater via thermohaline convection, stirred stratified fluids, geophysical fluids (i.e. potential vorticity (PV) staircases), Cahn-Hilliard Navier-Stokes fluids via phase separation and magnetized plasma, the subject of this volume. 
Inhomogeneous mixing has a bi-stable or multi-stable, and multi-scale character, which is intrinsic to the macroscopic layering, i.e. different regions mix unequally. 
This paper discusses inhomogeneous mixing and staircase structure in confined magnetized plasma. 

In tokamaks or stellarators, plasma is confined within nested magnetic flux surfaces formed by magnetic fields. 
The field lines lie on the flux surfaces.
The geometric structure of the field lines is characterized by the safety factor $q \approx \frac{rB_\phi}{RB_\theta}$ where $r$ and $R$ are minor and major radius and $B_\phi$ and $B_\theta$ are toroidal and poloidal magnetic field strength, respectively (Figure~\ref{fig:PD01}). 
It represents the ratio between the number of toroidal rotations and the number of poloidal rotations as one follows a field line. 
The flux surface of the rational $q$ value is called the rational surface on which the field lines are closed lines. 
For example, on the flux surface of $q=3/1$, a field line starting from any point will eventually return to its original position after completing three toroidal rotations and one poloidal rotation.

Turbulence in plasma (at low $\beta = (\mathrm{thermal~plasma~pressure/magnetic~pressure})$) consists of an ensemble of spatially localized convective cells.
They are often shaped to have the strong anisotropy ($k_\parallel / k_\perp \ll 1$ where $k_\parallel$ and $k_\perp$ are the parallel and perpendicular components of the wavenumber vector $\mathbf{k}$ with respect to the magnetic field $\mathbf{B}$) and are pinned to sites determined by the confining magnetic geometry.   
$\mathbf{k} \cdot \mathbf{B} = 0$ resonances or rational surfaces are one such pinning site. 
Typically, cells have radial correlation length $\lambda_\mathrm{corr}$ that is $\lambda_\mathrm{corr} / L_p \ll 1$. 
Here, $L_p  = \frac{\langle p \rangle}{d\langle p \rangle / dr}$ is a plasma (density or temperature) profile scale length. 
At the edge, $\lambda_\mathrm{corr}/L_p \le 1$. 
There are three obvious means by which mixing by plasma turbulence develops inhomogeneity.

\begin{figure}[!h]
%\centering\includegraphics[width=2.0in]{PD02_rev.png}
\centering\includegraphics[width=3.0in]{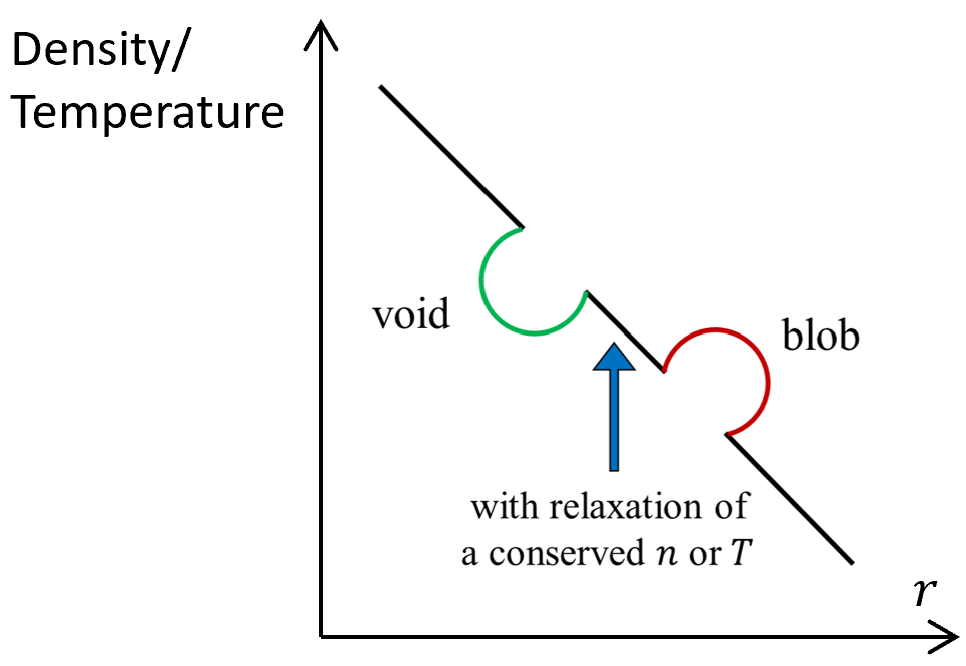}
\caption{Schematic diagram of birth of a blob-void pair due to local relaxation of gradients of a conserved plasma density \(n\) or temperature \(T\).}
\label{fig:PD02}
\end{figure}

\begin{enumerate}
\renewcommand{\labelenumi}{(\roman{enumi})}
\item Along with particles, heat, etc., the turbulence mixes plasma potential vorticity -- and thus vorticity -- so \(\langle {\tilde{v}}_{r}\nabla^{2}\tilde{\phi} \rangle \neq 0\) where $\tilde{v}_r$ and $\tilde{\phi}$ are fluctuations of radial plasma velocity and plasma potential, respectively.
Here, $\langle \rangle$ means averaging on the flux surface.
The Taylor identity then gives \(\langle {\tilde{v}}_{r}\nabla^{2}\tilde{\phi} \rangle = - \partial\langle {\tilde{v}}_{r}{\tilde{v}}_{\theta} \rangle/\partial r\) where $\tilde{v}_\theta$ is poloidal velocity fluctuation, so a non-zero Reynolds force develops, and drives sheared zonal flows on the scale of the spectral envelope.  
These zonal flows sink energy from the underlying relaxation process and thus impose a variable shearing pattern on the turbulence.
Consequently, mixing develops inhomogeneity.

\item The turbulence is interaction-driven, i.e. transfer of excitation energy from neighboring cells is comparable to, or exceeds, local energy input from gradients.
In this case, the production ratio for a given region, which can be written as \(PR = d(\langle {\tilde{v}}_{r}{\tilde{n}}^{2} \rangle)/\lbrack - \int_{}^{}{\langle {\tilde{v}}_{r}\tilde{n} \rangle(\partial\langle n \rangle/\partial r)dr\ }\rbrack\), is \(\geq\) 1. 
Here, $\tilde{n}$ is plasma density fluctuation, \(d(\langle {\tilde{v}}_{r}{\tilde{n}}^{2} \rangle)\) is the intensity flux differential across a region, and \(- \int_{}^{}{\langle {\tilde{v}}_{r}\tilde{n} \rangle(\partial\langle n \rangle/\partial r)dr\ }\) is the local production within it.
\(PR > 1\) indicates that a region is driven primarily by interaction, i.e. by turbulence spreading or avalanching from other regions, and not by local instability. 
Note that above definition of the production ratio is relevant to edge turbulence.  
Regarding core plasmas, ion temperature gradient (ITG) turbulence is a natural system for which to calculate a production ratio (a brief derivation of a production ratio for the gyrokinetic ITG model is given in Appendix for interested readers).
\(PR\) is related to, but not identical to, the more familiar Kubo number \(Ku\sim\tilde{v}\tau_\mathrm{corr}/\lambda_\mathrm{corr}\) where $\tau_\mathrm{corr}$ is the cell correlation time. 
\(Ku \geq 1\) indicates correlated, non-diffusive mixing. 
Recent experiments have measured large $PR$ and characterized its dependencies. 
Systems dominated by interactions tend to exhibit scale invariance across a range of mesoscopic scales $l$ bounded by the mean profile scale length ($L_p$) and the cell correlation length ($\lambda_\mathrm{corr}$), i.e. $\lambda_\mathrm{corr} < l < L_p$. Thus, transport phenomena can manifest on a broad range of scales, so the mixing likely is inhomogeneous.

\item Coherent structure production. 
Strong local relaxation of gradients of a conserved order parameter (i.e. density or temperature) gives birth to a blob-void pair (see Figure~\ref{fig:PD02}). 
Blobs -- particle excesses -- propagate down-gradient, while voids -- particle deficits -- propagate up-gradient.
These structures can enhance \(PR\), excite waves and turbulence, and induce a range of spatial couplings. 
The effective structure propagation distance (``mean free path'') defines a new, emergent, length scale in the problem.
\end{enumerate}

All of process (i)-(iii) will tend to induce inhomogeneous mixing.
The broad range of scales ($\lambda_\mathrm{corr} < l < L_p$) and complex nonlinear behavior of plasma turbulence make the study of inhomogeneous mixing one of the most challenging problems in plasma physics.
Section~\ref{sec:blob} discusses the microphysics of mixing in plasma, including turbulence properties, coherent structures and their effects on entrainment, spreading and avalanching, and zonal flows. 
Section~\ref{sec:staircase} introduces the mesoscopic consequences of inhomogeneous mixing in plasma, namely $\mathbf{E} \times \mathbf{B}$ staircases\footnote{$\mathbf{E}$ is the electric field. The $\mathbf{E} \times \mathbf{B}$ drift yields the poloidal flow in toroidal fusion plasmas via the radial electric field and toroidal magnetic field.}, or mesoscopic transport barriers. 
This section reports on a study of staircase structures and properties, and presents results from a multi-machine comparison. 
% Taken together, the results given in this paper paint a unique picture of the physics of layering in confined magnetized plasma, as discussed in Section~\ref{sec:sumdis}.
Taken together, Section~\ref{sec:sumdis} discusses what the recently improved understanding of mixing processes suggests for plasma staircase research. 

\section{Close-up of inhomogeneous mixing: blobs/voids in turbulence}
\label{sec:blob}

\subsection{Basic characteristics of blobs/voids}

\begin{figure}[!h]
%\centering\includegraphics[width=3.0in]{TL01_rev.png}
\centering\includegraphics[width=4.0in]{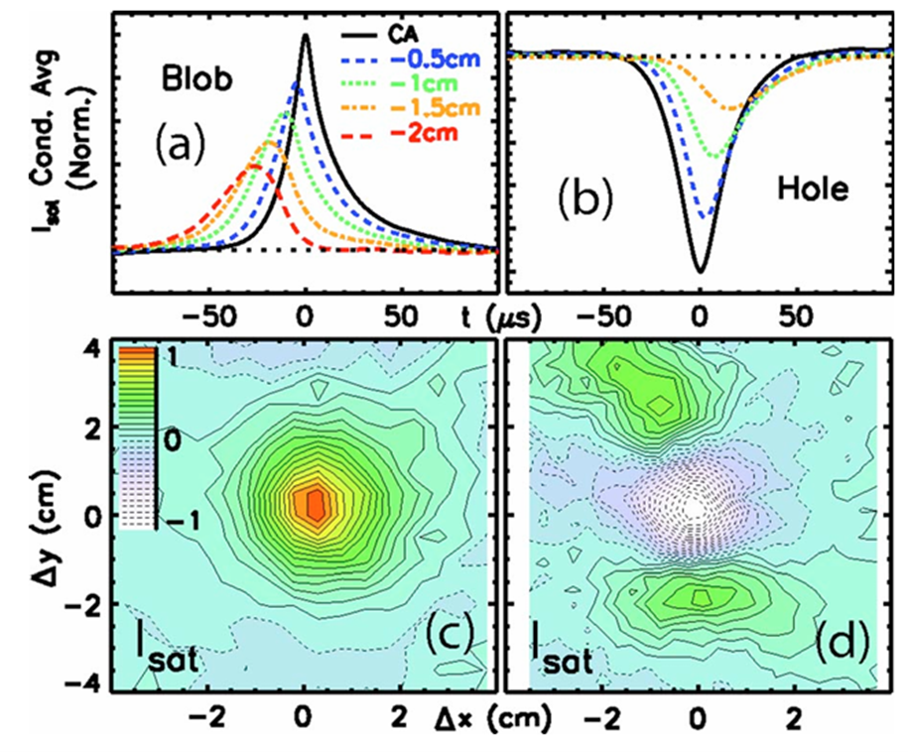}
\caption{Blob and void structures in the linear device LAPD \cite{carter2006}. In (a) and (b) are cross-conditional averages (CA) of ion saturation current $\tilde{I}_{sat}$ (representative of density fluctuations) for blob and void (hole) events, respectively, showing propagation of blob out of the plasma and void back into the plasma. Black line represents CA of $\tilde{I}_{sat}$ measurements by the reference (event detection) channel, and other lines represent CA of $\tilde{I}_{sat}$ measurements by channels at radially separated positions. For example, the red line is CA of $\tilde{I}_{sat}$ measurements by the channel which is 2~cm inward (closer to the plasma center) from the reference channel which is used to detect blob/void events. In (c) and (d) are two-dimensional cross-conditional averages to show the blob and void structures, respectively. $\Delta x$ and $\Delta y$ mean the relative measurement position against the reference channel position in the radial and poloidal direction, respectively. Reprinted from Carter et al. \cite{carter2006}. Copyright 2006 AIP.}
\label{fig:TL01}
\end{figure}

In the strongly turbulent edge region of toroidal/linear magnetically confined fusion plasmas, density or temperature fluctuations manifest intermittent/bursty behavior due to the presence of long-lived mesoscale coherent convective structures \cite{dippolito2011}. 
This kind of fluctuation structures/events, which exist within the ambient turbulence, are the so-called blobs and voids (or holes) \cite{dippolito2011, krasheninnikov2001, dippolito2002, boedo2003, dippolito2004, sladkomedova2023, long2024_structures, dippolito2008, CaoPRL2025, vianello2020, zweben2016, shesterikov2013}. 
Intermittent density fluctuations have been extensively observed and investigated in magnetic confinement fusion, while there are only few reports on intermittent temperature fluctuations \cite{kube2018, garcia2005}. 
Unless otherwise specified in the remainder of the article, blobs and voids refer to density blobs and density voids, respectively. 
The former are observed as magnetic-field-aligned filaments of excess density as compared with the background plasma, while the latter are observed as magnetic-field-aligned filaments of reduced density. 
Blobs and voids are born together to conserve density. 
In the two-dimensional cross-section normal to the magnetic field, these structures look just like ``blobs'' and ``voids'' \cite{dippolito2002, sladkomedova2023, carter2006, muller2007}.
Blob and void structures can be intuitively identified through the two-dimensional cross-conditional averages \cite{pecseli1989} of ion saturation current $\tilde{I}_{sat}$ (representative of density fluctuations), as shown by Figures~\ref{fig:TL01}(c) and \ref{fig:TL01}(d), respectively \cite{carter2006}.

Experimental studies show that blobs and voids can be observed at the boundary region of magnetic confinement plasma, either inside the last closed flux surface (LCFS) or outside the LCFS (in the scrape-off layer, SOL) on the low field side \cite{zweben2016, xu2009_jet, cheng2010, nold2010}. 
The positive/negative skewness of density fluctuations has been widely used to identify the local predominance of blobs/holes, respectively \cite{dippolito2011, dippolito2004}. 
Theoretical work indicates that the origin of blobs and voids is caused by instabilities such as curvature-driven interchange instability and drift wave instability \cite{myra2006a, russell2007, manz2015}. 
The former exhibits a strong intensity on the low magnetic field side with bad curvature in toroidal plasmas, while the latter can exist both in toroidal and linear plasmas.
The $\mathbf{E} \times \mathbf{B}$ sheared flow is also an important parameter controlling the blobs/voids' generation \cite{furno2008, agostini2007}. 
These explain why the blobs and voids are detected at the edge location of maximum normalized pressure gradient or inside the edge shear layer \cite{sladkomedova2023, cheng2013}. 
In the presence of a charge-dependent drift, e.g., curvature drift and grad-B drift, charge separation and the resulting $\mathbf{E} \times \mathbf{B}$ drift cause the blobs to move outward to the outer wall (down the mean gradient), while the voids propagate inward to the core (up the mean gradient) \cite{krasheninnikov2001}.
The opposite propagation directions of blobs and voids can be seen from the opposite time delays of cross-conditional averages (CA) of density fluctuations for blob and void events at different radial locations as shown in Figures~\ref{fig:TL01}(a) and \ref{fig:TL01}(b), respectively. 

Blobs and voids contribute significantly to the radial transport of energy and particle across the edge and SOL regions \cite{long2024_densitylimit, kotschenreuther2004, long2025_crossphase, ManzRMPP2025, labombard2004}. 
It should be noted that the core-boundary coupling is a key issue for future nuclear fusion reactors such as ITER and DEMO, that is, simultaneously achieving high-performance core plasma (to obtain high fusion power) and high-dissipation boundary plasma (to protect the plasma facing materials) \cite{ding2024, wang2021_detachment, long2024_poloidal}.
Therefore, it is necessary to study blobs and voids that are active in the plasma edge layers between the core and boundary. 
In recent years, new experimental research advancements and discoveries have provided novel insights and perspectives to understand the physical behavior and critical influence of blobs/voids. 
These include: (i) turbulence spreading induced by blobs/voids; (ii) turbulent layer broadening by blobs/voids; and (iii) interactions between blobs/voids and sheared flow, which will be discussed in detail below.

\subsection{Turbulence spreading induced by blobs/voids}
\label{sec:blobspread}
 
\begin{figure}[!h]
%\centering\includegraphics[width=4.0in]{TL02_rev.png}
\centering\includegraphics[width=5.0in]{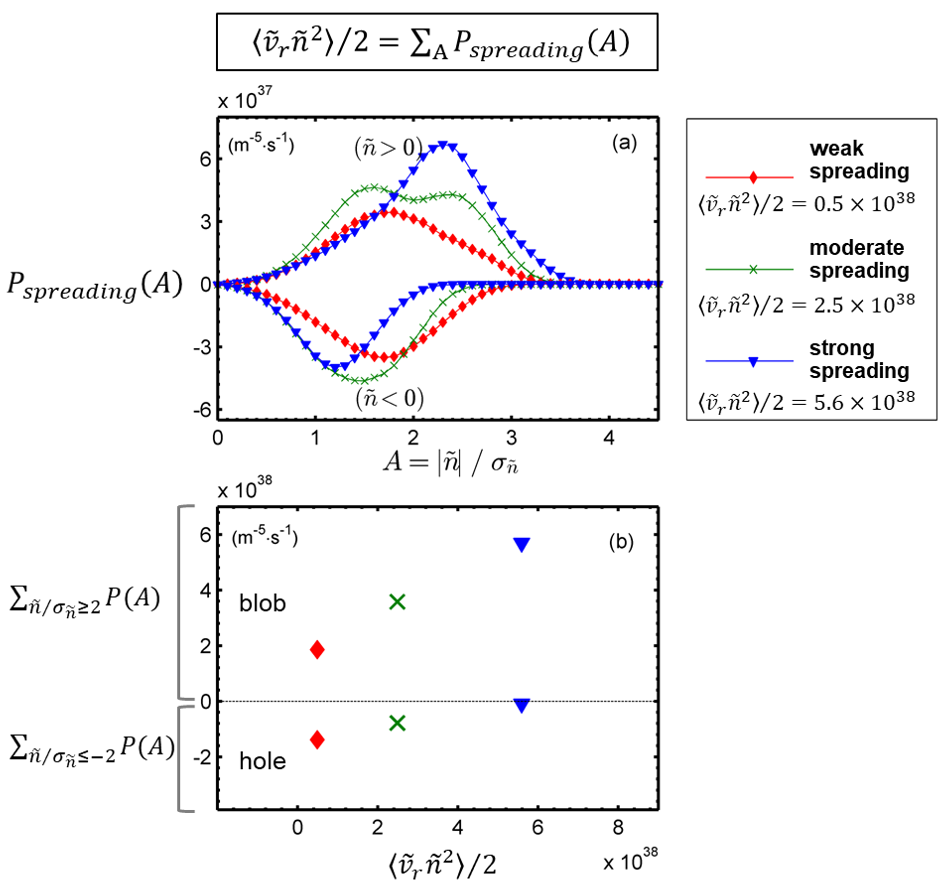}
\caption{(a) Turbulence spreading ``lip'' curves: distribution of turbulence spreading flux $P_{spreading}(A)$ relative to the density fluctuation amplitude $A = | \tilde{n} |/\sigma_{\tilde{n}}$; (b) contribution of blobs and voids to the total turbulence spreading $\langle {\tilde{v}}_{r}{\tilde{n}}^{2} \rangle/2$ in the plasma edge. Reprinted from Long et al. \cite{long2024_structures}. Copyright 2024 IOP.}
\label{fig:TL02}
\end{figure}

Turbulence spreading means the propagation of turbulence intensity or energy in space.
This is often quantified as the flux of turbulence intensity or energy across the magnetic field. 
For example, the turbulence spreading flux for density turbulence intensity is calculated as $\langle {\tilde{v}}_{r}{\tilde{n}}^{2} \rangle/2$ \cite{manz2015, gurcan2006, HahmJKPS2018}. 
Spreading of turbulence occurs via relaxation of inhomogeneous turbulence and entrainment of laminar, or more weakly turbulent regions \cite{gurcanPRL2006}. 
One well known example is an expanding wake formed downstream of a moving object, which corresponds to turbulence spreading from the fully turbulent core \cite{townsend1949}.
The turbulence intensity field can be decoupled from the local instability growth rate due to turbulence spreading.
In-depth experimental research of the relation of turbulence spreading to blobs-voids in edge plasmas via Langmuir probe measurements on J-TEXT tokamak has been reported recently \cite{long2024_structures}. 
It is found that the turbulence spreading flux originating from positive density fluctuations is positive, meaning the outward propagation of the positive $\tilde{n}$ turbulence intensity. 
The turbulence spreading flux originating from negative density fluctuations is negative, meaning the inward propagation of the negative $\tilde{n}$ turbulence intensity. 
This is shown by the turbulence spreading ``lip'' curves in Figure~\ref{fig:TL02}(a), which presents the distribution of turbulence spreading flux $P_{spreading}(A) = \sum_{A - 0.05 < |\tilde{n}|/\sigma_{\tilde{n}} < A + 0.05} (\tilde{v}_r \tilde{n}^2 /2) / M$ calculated using either positive or negative $\tilde{n}$ where $A = |\tilde{n}|/\sigma_{\tilde{n}}$ and $\sigma_{\tilde{n}}$ is the standard deviation of plasma density fluctuation \cite{long2024_structures}. 
$M = 6000$ indicates the total number of sampling points of $\tilde{n}$ and $\tilde{v}_r$. 
The distribution curve of turbulent spreading flux originating from positive density fluctuations is always positive (corresponding to outward spreading), i.e. $P_{spreading}(A) > 0$ for $\tilde{n}/\sigma_{\tilde{n}} > 0$, which outlines the upper edge of the ``lip''. 
The distribution curve of turbulent spreading flux originating from negative density fluctuations is always negative (corresponding to inward spreading), i.e. $P_{spreading}(A) < 0$ for $\tilde{n}/\sigma_{\tilde{n}} < 0$, which outlines the lower edge of the lip. 
The inward turbulence spreading caused by voids has also been further demonstrated by the experimental measurements via beam emission spectroscopy (BES) on the DIII-D tokamak \cite{khabanov2024} and is the subject of a newly developed theoretical model \cite{CaoPRL2025}.

The net total turbulence spreading flux $\langle {\tilde{v}}_{r}{\tilde{n}}^{2} \rangle/2$ is taken as the sum of positive and negative components, and the weak, moderate, and strong spreading flux cases are shown in Figure~\ref{fig:TL02}(a). 
The distribution function of turbulence spreading flux is almost symmetric for the weak spreading case (red diamonds). 
The distribution function becomes more strongly asymmetric (green crosses and blue inverted triangles) for the stronger net outward spreading flux cases.  
As shown in Figure~\ref{fig:TL02}(b), while the contribution by the negative high-amplitude density fluctuations to the net flux decreases, the contribution by the positive high-amplitude density fluctuations increases, which is especially true for contributions with absolute amplitudes higher than $2\sigma_{\tilde{n}}$. 
The outward turbulence spreading is caused by the blobs-dominant events with $\tilde{n} \geq 2\sigma_{\tilde{n}}$, and the inward spreading is caused by the holes-dominant events with $\tilde{n} \leq - 2\sigma_{\tilde{n}}$. 
This blob-void asymmetry emerges as crucial to turbulence spreading.

The result above highlights the importance of blob/void convective structures in conveying the spreading of turbulence in the edge plasmas, and brings about a challenge to diffusive models of turbulence spreading in the form of Fick's law as often assumed and long used \cite{long2024_structures, HahmJKPS2018}.
Also, turbulence spreading is closely related to avalanche transport \cite{HahmJKPS2018} in core plasmas.
The mechanism of turbulence spreading described above is expected to be critical in understanding the avalanche transport phenomena observed in core plasmas \cite{PolitzerPRL2000, PanNF2015, Choi:2019wy, KinSR2023, hongNF2026} as well.

\subsection{Turbulent layer broadening by blobs/voids}

Edge plasmas are usually strongly turbulent in linear plasma devices and toroidal fusion devices \cite{xu2010_fourier, yan2019, sun2021, sladkomedova2023}. 
Since blobs propagate outward and detach from the bulk plasma, while voids propagate inward and so stir the core plasma, it is natural to consider the roles of blobs/voids in the broadening of turbulent layer.

Regarding the turbulent layer broadening by blobs, our recent experimental study shows that in high collisionality regime, the enhanced turbulence spreading induced by blobs can effectively broaden the power scrape-off width $\lambda_{q_{\parallel}}$ \cite{long2024_conf}. 
\(\lambda_{q_{\parallel}}\) is one of the most important practical quantities of the SOL since it controls the plasma power exhaust and thus governs the problems of heat removal and melting of plasma facing components (PFCs) \cite{stangeby2010}.
The SOL collisionality parameter is used to distinguish the regimes of the SOL, including the sheath-limited regime with low $\nu_{SOL,e}^{*} \approx 10^{-16} \frac{n_{LCFS}L_\parallel}{T_{\mathrm{e},LCFS}^2}$ and the conduction-limited regime with high $\nu_{SOL,e}^{*}$ \cite{stangeby2000}.
Here, $n_{LCFS}$ is the plasma density at the last closed flux surface (LCFS), $T_{e,LCFS}$ is the electron temperature at the LCFS, and $L_\parallel$ is the connection length of magnetic field lines. 
Probe measurements of two discharges with different toroidal magnetic field $B_{\phi}$, plasma current $I_{p}$ and central line-averaged density $\overline{n}$ on J-TEXT are compared. 
In all these discharges, $\nu_{SOL,e}^{*}$ is 20-200, which is much larger than 10. 
Thus, the SOL is in the conduction-limited regime. 
For this regime, the power scrape-off width or power decay length $\lambda_{q_{\parallel}} \approx 2/7\lambda_{T}$.
 Here, $\lambda_{T}$ is the temperature decay length.

It is constructive to compare the experimental widths $\lambda_{q,EXP}$ with the theoretical predictions from neoclassical heuristic drift model and turbulent broadening model.
For neoclassical heuristic drift (HD) model, note that the preliminary HD (PHD) model is applied for zero collisionality \cite{goldston2011}, while the generalized HD (GHD) model is applied for finite collisionality \cite{brown2021}. 
For turbulent broadening (TB) model, the impact of turbulence spreading on SOL broadening is considered \cite{long2024_structures, chu2022}. 
Turbulence intensity $E$ sets the power scrape-off layer width beyond the PHD width as $\lambda_{q,TB} = ( \lambda_{q,PHD}^{2} + {\tau_{\parallel}}^{2}E )^{1/2}$.
Here, $\tau_{\parallel} = L_{\parallel}/c_{s}$ is the SOL residence time of heat where $c_s=(T_\mathrm{e}/m_\mathrm{i})^{1/2}$ is the ion sound speed. 
As an estimate, the turbulence intensity is given by $E = V_{I}^{2}\tau_{ac}^{2}/\tau_{\parallel}^{2}$ \cite{long2024_structures}. 
Here, ``mean jet velocity'' $V_{I}$ represents the velocity of turbulence spreading \cite{townsend1949}, and $\tau_{ac}$ is the auto-correlation time of density fluctuations.

{\renewcommand{\arraystretch}{1.1}
\begin{table}
\caption{Correlation between blobs' dynamics and power scrape-off widths $\lambda_q$}%%%Table caption goes here
\label{table:BLOB}
\resizebox{\textwidth}{!}{
\begin{tabular}{p{1cm} p{1cm} p{1.5cm} p{1.5cm} p{1.5cm} p{1.5cm} p{1.5cm} p{1.5cm} p{1.5cm} p{1cm} p{1cm} p{1cm} p{1cm}}%%%The number of columns has to be defined here
\hline
$B_\phi$ [T]    & $I_p$ [kA] & $\overline{n}$ [$10^{19}~m^{-3}$] & SOL collisionality $\nu_{SOL,e}^{*}$ & Turbulence spreading $\langle {\tilde{v}}_{r}{\tilde{n}}^{2} \rangle/2$ & Blobs burst rate $\gamma_b$ [kHz] & Blobs lifetime $\tau_{life}$ [$\mu$s] & Blobs speed $v_{prop}$ [m/s] & Blobs characteristic length $l_r$ [mm] & $\lambda_{q,PHD}$ [cm] & $\lambda_{q,GHD}$ [cm] & $\lambda_{q,TB}$ [cm] & $\lambda_{q,EXP}$ [cm] \\
\hline
2.2 & 187 & 4.1 & 33 (lower)  & 0.5$\times 10^{38}$ & 7.9 & 8.1 & 70  & 0.6 & 0.175 & 0.370 & 0.181 & 0.205   \\
1.6 & 126 & 3.9 & 92 (higher) & 5.6$\times 10^{38}$ & 12  & 11  & 630 & 6.6 & 0.194 & 1.127 & 0.350 & 0.341   \\
\hline
\end{tabular}
}
\vspace*{-4pt}
\end{table}%%%End of the table
}

Table~\ref{table:BLOB} presents the experimental measured power scrape-off widths $\lambda_{q,EXP}$ and theoretical predicted widths for the two discharges with different blobs' dynamics. 
The temporal and spatial features of blobs, including burst rate $\gamma_{b}$, lifetime $\tau_{life}$, radial propagation speed $v_{prop}$ and characteristic radial length $l_{r}$, are obtained via the routine conditional average methods \cite{long2024_structures, cheng2010, pecseli1989}.
Compared with the case of lower SOL collisionality $\nu_{SOL,e}^{*} \sim 33$ with weaker turbulence spreading, in the case of higher SOL collisionality $\nu_{SOL,e}^{*} \sim 92$ with stronger turbulence spreading, the blobs' burst rate $\gamma_{b}$ increases by 1.5 times, lifetime $\tau_{life}$ increases by 1.4 times, radial propagation speed $v_{prop}$ increases by 9.0 times, characteristic radial length $l_{r}$ increases by 1.1 times, and the experimental measured power scrape-off widths $\lambda_{q,EXP}$ correspondingly increases by 1.7 times. 
Note that for the case of higher $\nu_{SOL,e}^{*}$ with more active blobs, the preliminary heuristic drift model ($\lambda_{q,PHD}\sim 0.194\ cm$) obviously underestimates power scrape-off width ($\lambda_{q,EXP}\sim 0.341\ cm$) and the generalized heuristic drift (GHD) model ($\lambda_{q,GHD}\sim 1.127\ cm$) significantly overestimates power scrape-off width, while the turbulence broadening model ($\lambda_{q,TB}\sim 0.350\ cm$) is in good agreement with the experimental measurement. 
The result above suggests the important roles of blobs and turbulence spreading in the broadening of power scrape-off width in the high collisionality regime. 
This is also consistent with previous experimental and simulation work \cite{wu2021_hl2a_sol, li2019_iter, wu2023_heatflux}.

Recently, a new theoretical model has demonstrated that the heretofore ignored process of void emission can drive a broad turbulent layer of width $\sim 100 \rho_{s}$ for typical parameters ($\rho_{s}$ is the ion sound speed gyroradius) \cite{CaoPRL2025}. 
Since the voids are closely related to inward turbulence spreading, this once again evokes the previous relevant suggestion that models of the pedestal structure or transport barrier should include the influence of coherent structures and turbulence spreading \cite{singh2020}. 
More experimental validation of this prediction should be carried out.

\subsection{Interactions between blobs/voids and zonal flow}
\label{sec:blobflow}

The interactions between blobs/voids and sheared flow are complex. 
On the one hand, the $\mathbf{E} \times \mathbf{B}$ sheared flow can cause the decorrelation, distortion, and breaking of blobs/voids, and regulate the formation of blobs/voids and their transport \cite{shesterikov2012_textor, biglari1990}. 
Recent experimental studies indicate that improvement of particle confinement and higher operational density are achieved in positive biasing discharges of J-TEXT as the outward motion of the blobs is blocked, and the turbulent transport is reduced by stronger externally driven sheared flow \cite{li2022_jtext, ke2022_electrode}. 
In the experiments which scan line-averaged density and plasma current approaching the Greenwald density limit, the turbulence spreading induced by blobs is found to decrease as the $\mathbf{E} \times \mathbf{B}$ shearing rate increases, while the turbulence spreading induced by voids is insensitive to $\mathbf{E} \times \mathbf{B}$ shearing rate \cite{long2024_densitylimit}.

\begin{figure}[!h]
%\centering\includegraphics[width=2.0in]{TL03_rev.png}
\centering\includegraphics[width=3.0in]{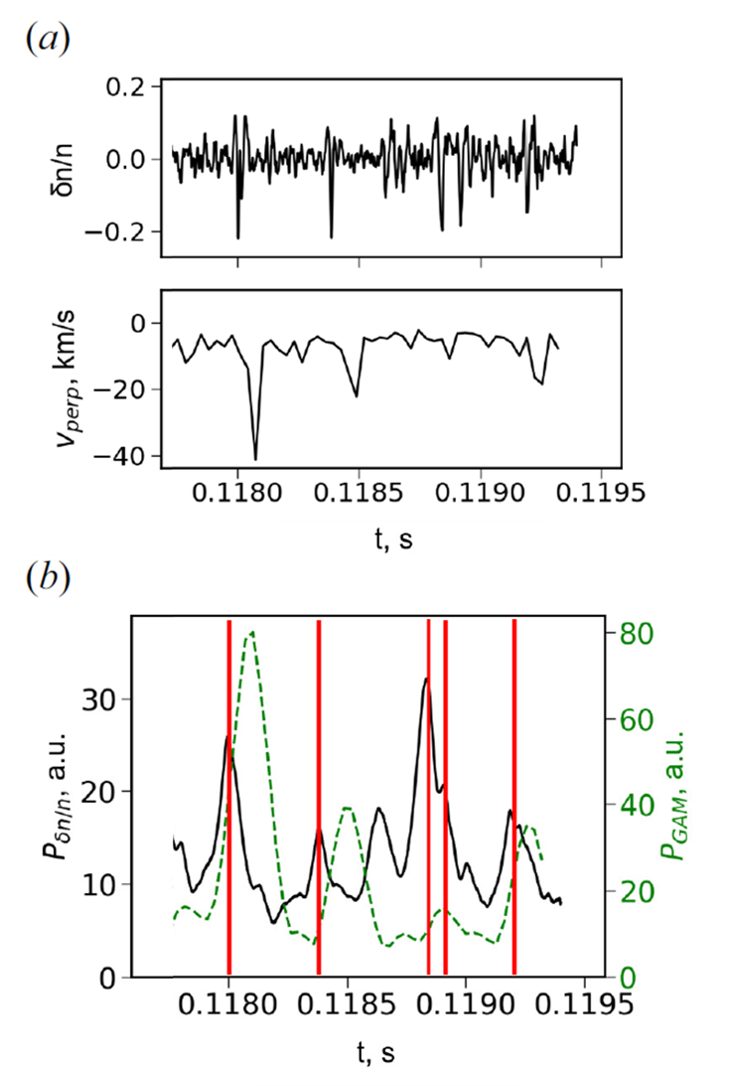}
\caption{(a) Time series of the relative density fluctuations $\delta n/n$ (top) and of the perpendicular velocity (bottom); (b) wavelet power of $\delta n/n$ (black solid line) and of the poloidal velocity fluctuations near the GAM frequency (green dashed line) in the plasma edge. Vertical red lines denote the times when voids are present. Reprinted from Sladkomedova et al. \cite{sladkomedova2023}. Copyright 2023 Cambridge University Press.}
\label{fig:TL03}
\end{figure}

On the other hand, radially propagating blobs/voids can excite drift waves and drive intermittent zonal flow \cite{CaoPRL2025, ivanov2020}.
This result could qualitatively explain the strong correlation between zonal flow power bursts and the detection of voids measured by BES diagnostic in experiments on MAST spherical tokamak \cite{sladkomedova2023}. 
As shown in Figure~\ref{fig:TL03}(a), shortly after the voids of the relative density fluctuations $\delta n/n$ emerge, the perpendicular velocity $v_{perp}$ (approximate to poloidal velocity $v_{\theta}$) exhibits a similar intermittent bursty behavior. 
A sharp peak in the cross correlation and a near zero phase of the cross spectra at $\sim 8$~kHz between two poloidally separated $v_{perp}$ channels suggest the existence of the poloidally symmetric coherent flows. 
These are called geodesic acoustic modes (GAMs), which are a branch of zonal flow with finite frequency \cite{DiamondPPCF2005}. 
Gas puff imaging (GPI) measurements on the HL-2A tokamak, which track the blob motions, suggest that the frequency of blob/void structures are more likely to be higher than $10$~kHz \cite{yuan2019}.

In Figure~\ref{fig:TL03}(b), the black solid curve shows the time series of the wavelet power of $\delta n/n$ ($P_{\delta n/n}$) integrated over the frequency range of 10--500~kHz, the green dashed curve shows the time series of the wavelet power of the perpendicular velocity fluctuations near the GAM frequency integrated over 6--9~kHz ($P_{GAM}$), and the vertical red lines denote the times when voids are present. 
It is found that the burst of density voids occurred almost simultaneously with the transient rise of the broadband turbulent fluctuation power $P_{\delta n/n}$. 
It is also evident that a transient rise of GAM power $P_{GAM}$ occurs $\sim 1$~ms after the burst of density voids.
Besides, recent experimental observations in the plasma edge of TJ-K stellarator also support this physical mechanism \cite{dumeratPPCF2025}. 
Blobs in the SOL are found to cause potential fluctuations in a zonal band inside the separatrix in a unidirectional causality by using probe measurements and convergent cross mapping methods. 
These results indicate that blobs and voids can play an important role in driving intermittent zonal flow.
Moreover, formation of such void or blob driven zonal flows can initiate staircase formation, linking  microphysics of blob/void and mesoscopic structure formation. 
It is to staircases and other mesoscopic phenomena that we turn in the following section.

\section{Mesoscopic structure by inhomogeneous mixing: the $\mathbf{E} \times \mathbf{B}$ staircase}
\label{sec:staircase}
\subsection{Introduction to the $\mathbf{E} \times \mathbf{B}$ staircase}

\begin{figure}[!h]
%\centering\includegraphics[width=4.5in]{GYSELA.png}
\centering\includegraphics[width=5.5in]{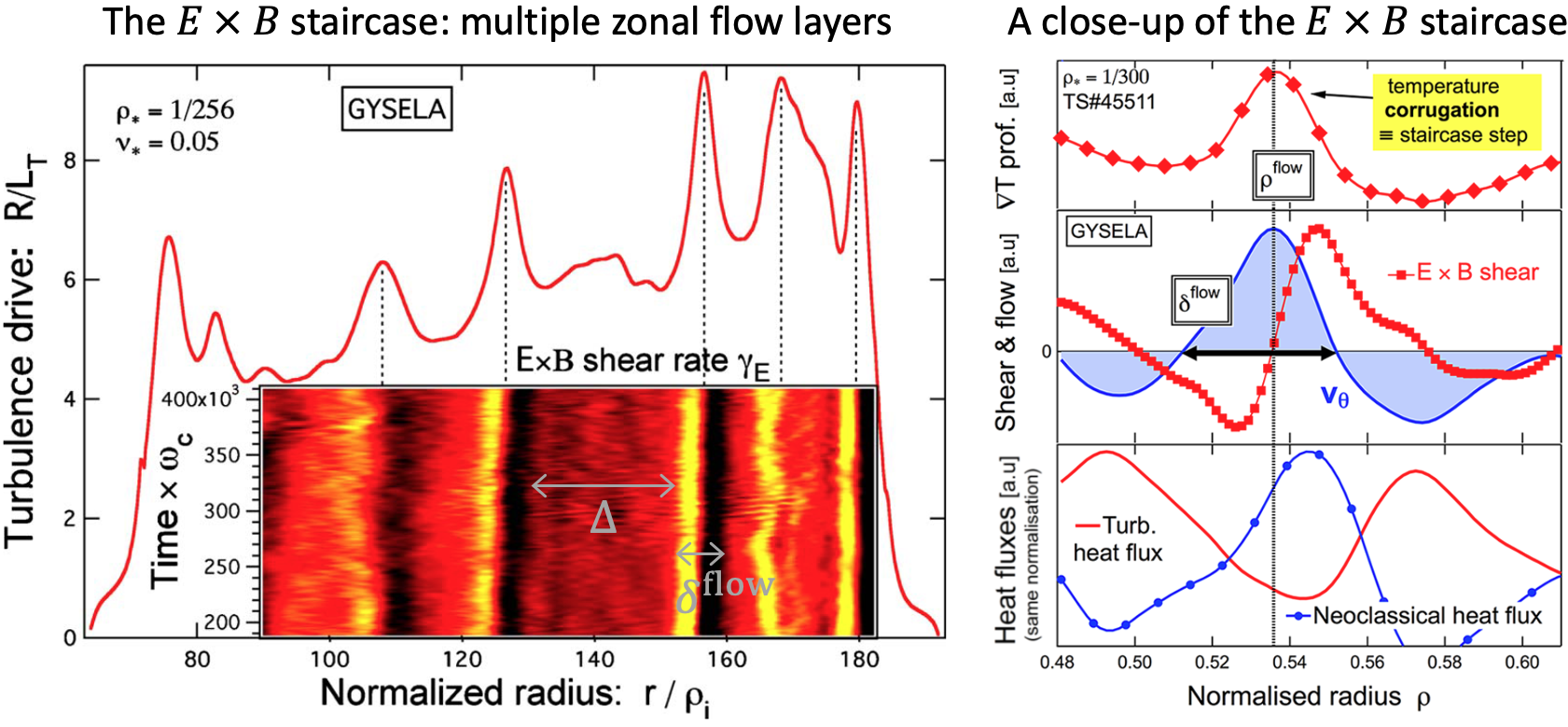}
\caption{The $\mathbf{E} \times \mathbf{B}$ staircase observed in the GYSELA simulation with the ion collisionality $\nu_*=0.05$ and the normalized ion gyroradius $\rho_* = \rho_\mathrm{i}/a = 1/256$. Reprinted from Dif-Pradalier et al. \cite{DifPRE2010} (Copyright 2010 APS) and Dif-Pradalier et al. \cite{DifPRL2015} (Copyright 2015 APS).}
\label{fig:GYSELA}
\end{figure}

The $\mathbf{E} \times \mathbf{B}$ staircase refers to a self-organized $\mathbf{E} \times \mathbf{B}$ zonal flow layers, separated by a mesoscopic scale.
These sheared zonal flow layers act as weak or permeable transport barriers, resulting in a corrugated pressure profile resembling a staircase.
The treads and riser of a corrugated profile correspond to the strong and weak mixing regions, respectively, representing an inhomogeneous mixing \cite{DritschelJAS2008} of pressure. 
It was first identified in the full-f flux-driven electrostatic and adiabatic electron gyrokinetic simulations using GYSELA and XGC1 \cite{DifPRE2010, DifNF2017}. 
Figure~\ref{fig:GYSELA} shows the normalized temperature gradient profile whose peaks align with the locations of strong flow shear layers. 
The $\mathbf{E} \times \mathbf{B}$  staircase can be most generally characterized by two spatial scales $\delta^\mathrm{flow}$ and $\Delta$. 
$\delta^\mathrm{flow}$ indicates the radial extent of each zonal flow layer, and $\Delta$ (tread width) is the distance between adjacent zonal flow layers or the radial extent of the mixing zone.
These two may not be clearly distinguishable in some cases though \cite{WangNF2018}. 

The $\mathbf{E} \times \mathbf{B}$ staircase is a pattern which spontaneously emerges in the near-marginal regime \cite{DifNF2017}. 
The near-marginal regime means a state of the plasma which fluctuates close to the instability threshold.
The near-marginal state is maintained by transport processes. 
The turbulence transport in this regime is distinct from the expectations of commonly used reduced models based on the local diffusive paradigm \cite{DiamondPoP1995, GillotPPCF2023}. 
In particular, large Kubo number $Ku > 1$  is predicted in a near-marginal regime \cite{GillotPPCF2023}. 
%The Kubo number is defined as the ratio between the distance traveled by a particle with typical velocity $\tilde{v}$ for the correlation time $\tau_\mathrm{corr}$ of an ambient potential eddy to the correlation length $\lambda_\mathrm{corr}$ of the eddy 
The Kubo number \cite{VladRRP2008} can be written as the ratio of the correlation time $\tau_\mathrm{corr}$ to the eddy turnover time $\tau_\mathrm{turn} = \lambda_\mathrm{corr}/\tilde{v}$.
When $Ku < 1$, which is the usual criterion for validity of a quasilinear theory, the transport is more like diffusion by random kicks by the rapidly decorrelating potential eddies. 
When $Ku > 1$, the particles would have sufficient time to interact with the potential eddies and their transport can exhibit other behaviors including either a sub-diffusion by being trapped within a localized eddy or a super-diffusion by traveling along a streamer \cite{DiamondNF2001}, or the radially successively-spreading \cite{HahmJKPS2018, long2024_structures} or phase-matching \cite{KishimotoPTRA2023} eddies. 
The latter can appear as avalanches, or ballistic propagation of (density or temperature) blobs and voids whose size distribution follows the self-similar or scale-invariant power-law \cite{PolitzerPRL2000, PanNF2015, Choi:2019wy, KinSR2023}, when a multitude of metastable states exists as in the self-organized criticality (SOC) system \cite{DiamondPoP1995, HahmJKPS2018, SanchezPPCF2015}. 
The $\mathbf{E} \times \mathbf{B}$ staircase forms as avalanches are regulated within $\Delta$ of spontaneously emerging zonal flow layers. 

The co-existence of avalanches and zonal flow layers makes the turbulence transport in the near-marginal regime more complicated.
Understanding the $\mathbf{E} \times \mathbf{B}$ staircase would enable closing the gap between self-consistent -- but expensive -- gyrokinetic simulations and reduced -- but practical -- models \cite{GillotPPCF2023}. 
The rest of this Section~\ref{sec:staircase} presents a brief summary of rare experimental observations and characterizations of the $\mathbf{E} \times \mathbf{B}$ staircase.
The aim is to enhance its understanding and to encourage further experimental research. 

\subsection{Observations of the $\mathbf{E} \times \mathbf{B}$ staircase}

In a broad sense, the $\mathbf{E} \times \mathbf{B}$ staircase is an array of weak transport barriers formed by localized multiple shear flow layers extending through the whole plasma. 
The conventional methods to identify a macroscopic transport barrier in experiments are based on one dimensional profiles of the measured pressure or the estimated radial electric field. 
However, identifying weak transport barriers only with one dimensional measurement of profiles may be challenging.
The weak profile corrugation would not be clearly distinguished from measurement noise. 
To overcome this limitation, different approaches have been implemented in Tore Supra \cite{DifPRL2015, HornungNF2017}, HL-2A \cite{LiuPoP2021}, DIII-D \cite{Ashourvan:2019ek}, KSTAR \cite{Choi:2019wy, ChoiPPCF2024}, TJ-II \cite{MilligenNF2017}, and W7-X \cite{MilligenNF2018} plasmas as described below. 
The experimental conditions are summarized in Table~\ref{table:STEX}. 
While the experimental findings may depend on a specific device configuration and plasma scenario, they might also reveal different aspects of the $\mathbf{E} \times \mathbf{B}$ staircase. 

{\renewcommand{\arraystretch}{1.1}
\begin{table}
\caption{Experiments on mesoscopic transport barriers. Key plasma parameters and experiment condition are listed: plasma major radius $R_0$, minor radius $a$, toroidal magnetic field strength (at $R_0$) $B_\phi$, plasma current $I_p$, heating methods (radio frequency (RF) resonance heating; neutral beam injection (NBI); electron cyclotron wave resonance heating (ECH)), dominant micro-instabilities (ion temperature gradient mode (ITG) with $\eta_\mathrm{i} \equiv L_n / L_{T_\mathrm{i}}$ drive where $L_n$ and $L_{T_\mathrm{i}}$ are the density and ion temperature gradient scale lengths, respectively; trapped electron mode (TEM); electron temperature gradient mode (ETG)), the ion collisionality $\nu_*$, and the inverse normalized ion gyroradius $\rho_*^{-1} = a/\rho_\mathrm{i}$.}
\label{table:STEX}
\resizebox{\textwidth}{!}{
\begin{tabular}{llllllllll}%%%The number of columns has to be defined here
\hline
Device     & $R_0$ [m] & $a/R$ & $B_\phi$ [T] & $I_p$ [MA] & Heating [MW] & Instability & $\nu_*$ & $\rho_*^{-1}$   \\
\hline
Tore Supra \cite{DifPRL2015, HornungNF2017} & 2.39 & 0.31 & 2.80 & 0.80 & Ohmic or RF $\sim$ 3          & ITG $\eta_\mathrm{i} \sim$ 2--3   & 0.01--4    & 200--500   \\
HL-2A \cite{LiuPoP2021}                     & 1.65 & 0.24 & 1.30 & 0.16 & NBI $\sim$ 0.74               &                                   &            & $\sim$110  \\
DIII-D \cite{Ashourvan:2019ek}              & 1.68 & 0.36 & 1.90 & 1.15 & NBI $\sim$ 10, ECH $\sim$ 3.4 & TEM/ITG, ETG                      & 0.05--0.15 & $\sim$150  \\
KSTAR \cite{Choi:2019wy, ChoiPPCF2024}      & 1.80 & 0.27 & 3.00 & 0.50 & NBI $\sim$ 4, ECH $\le$ 1     & ITG $\eta_\mathrm{i} \sim$ 1.5--3 & 0.15--0.3  & $\sim$200  \\
TJ-II \cite{MilligenNF2017}                 & 1.50 & 0.13 & 0.95 & N/A  & ECH $\sim$ 0.5                & Resistive MHD                     &            & $\sim$65   \\
W7-X \cite{MilligenNF2018}                  & 5.50 & 0.09 & 3.00 & N/A  & ECH $\sim$ 0.6--2.0           & Resistive MHD                     &            & $\sim$230  \\
\hline
\end{tabular}
}
\vspace*{-4pt}
\end{table}%%%End of the table
}

\begin{figure}
%\centering\includegraphics[width=4.5in]{TSHL.png}
\centering\includegraphics[width=5.5in]{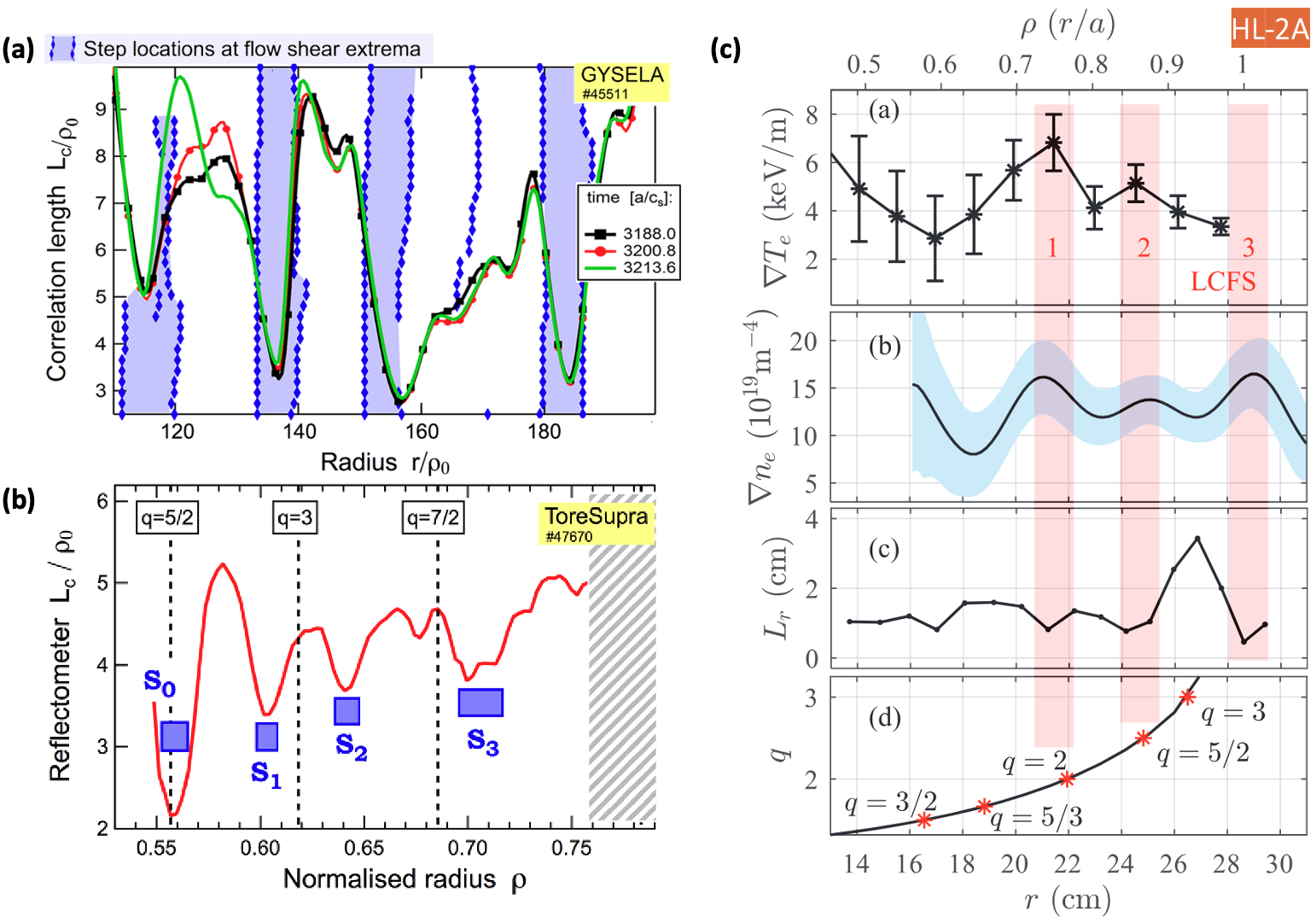}
\caption{(a) The reduction of the radial correlation length near the shear flow extrema in the GYSELA simulation. Black, red, and green lines represent the radial profile of the correlation length normalized by the ion gyroradius, $L_c / \rho_0$, at different normalized times $a/c_s$ where $a$ is the minor radius and $c_s=(T_\mathrm{e}/m_\mathrm{i})^{1/2}$ is the ion sound speed. (b) The measured correlation length profile, showing multiple minima locations expecting the shear flow layers ($S_0$--$S_3$) in the Tore Supra experiment. Reprinted from Dif-Pradalier et al. \cite{DifPRL2015}. Copyright 2015 APS. (c) The radial profiles of the temperature and density gradients, the radial correlation length, and the safety factor $q$ in the HL-2A experiment. Reprinted from Liu et al. \cite{LiuPoP2021}. Copyright 2021 AIP.}
\label{fig:TSHL}
\end{figure}

The first approach to identify the $\mathbf{E} \times \mathbf{B}$ staircase in experiments is to measure the variation of turbulence correlation length or the tilt of turbulent structures by the shear flow layers. 
For example, the local minima of the turbulence correlation length measurements along the radius were identified in Tore Supra L-mode plasmas \cite{DifPRL2015, HornungNF2017}. 
It was shown using the GYSELA data that the flow shear extrema are well correlated with the correlation length minima as shown in Figure~\ref{fig:TSHL}(a). 
Due to a possible meandering behavior of the $\mathbf{E} \times \mathbf{B} $ staircase \cite{DifPRL2015}, the fast-sweeping X-mode reflectometry on Tore Supra was used to get an almost instantaneous radial measurements of turbulent fluctuations to obtain the correlation length profile. 
Figure~\ref{fig:TSHL}(b) shows that multiple local minima ($S_0 $—$S_3$) were identified, and they are not closely associated with the rational $q$ surfaces except $S_0$. 
Note that measurements were done in plasmas without MHD activity to avoid unwanted effects in the analyses. 
The reversed tilt angle of the time-radius correlation function was also provided as further evidence of the shear flow layers \cite{HornungNF2017}. 

In HL-2A L-mode plasmas \cite{LiuPoP2021}, beam emission spectroscopy was used to measure the radial variation of the turbulence correlation length and also to investigate the tilt of density fluctuations. 
Measurements were done in a relatively quiet phase in between MHD associated bursts. 
Several local minima of the correlation length were identified as shown in Figure~\ref{fig:TSHL}(c) and two of them seem to be correlated with the low order rational surfaces. 
The corresponding changes in the radial-poloidal wavenumber spectrum of density fluctuations (more scattered in the radial wavenumber space) were observed around the expected shear flow layer locations \cite{LiuPoP2021}.

The second approach is to improve the credibility of the measured radial corrugation in plasma profiles using additional information from multi transport channels, fluctuations, or two-dimensional data. 
Figure~\ref{fig:TSHL}(c) shows measurements of both electron temperature and density gradients, exhibiting multiple peaks aligned with the expected location of the shear flow layers from the turbulence analysis in the HL-2A plasmas \cite{LiuPoP2021}. 
A region of relatively enhanced mixing (more efficient turbulence spreading), which corresponds to a larger correlation length region in Figure~\ref{fig:TSHL}(c), was identified in between the peaks (the transport barriers or suppressed mixing regions) of plasma profiles. 

\begin{figure}
%\centering\includegraphics[width=5.0in]{D3KS.png}
\centering\includegraphics[width=6.0in]{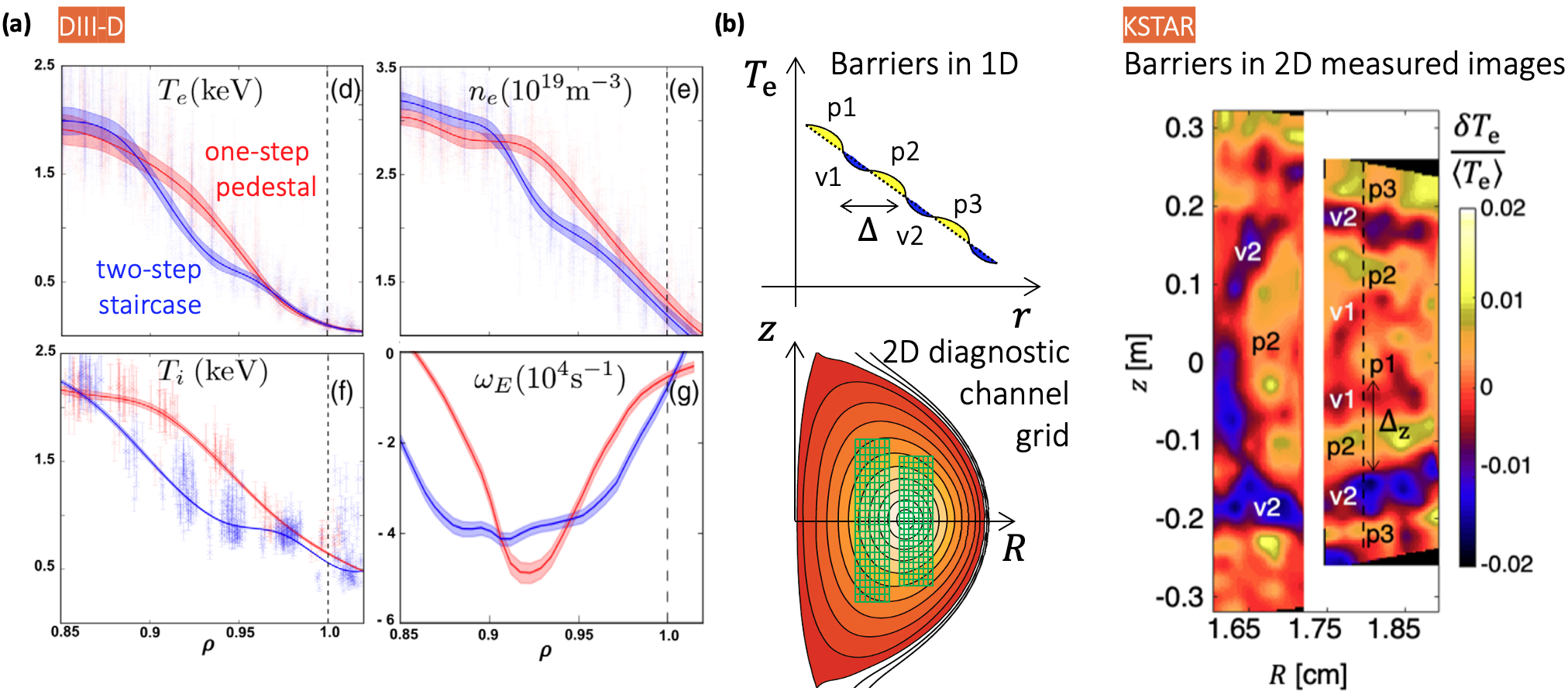}
\caption{(a) The radial profiles of the electron temperature $T_\mathrm{e}$ and ion temperature $T_\mathrm{i}$, density $n_\mathrm{e}$, and $\mathbf{E} \times \mathbf{B}$ rotation frequency in the DIII-D experiment. Reprinted from Ashourvan et al. \cite{Ashourvan:2019ek}. Copyright 2019 APS. (b) The two-dimensional (2D) measurement of the radial corrugation of normalized electron temperature variation $\frac{\delta T_\mathrm{e}}{\langle T_\mathrm{e} \rangle} = 
\frac{T_\mathrm{e} - \langle T_\mathrm{e} \rangle}{\langle T_\mathrm{e} \rangle}$ in the KSTAR experiment. Here, $\langle \rangle$ means the time average. Peaks (p1--p3) and valleys (v1--v2) appear as concentric jet-like patterns in 2D. $\Delta$ ($\Delta_z$) indicates the tread width of temperature staircase along the radial (vertical) direction. Reprinted from Choi et al. \cite{Choi:2019wy}. Copyright 2019 IOP.}
\label{fig:D3KS}
\end{figure}

In DIII-D H-mode edge plasmas \cite{Ashourvan:2019ek}, the axisymmetric two-step profiles were observed in multi channels such as density and electron and ion temperatures as shown in Figure~\ref{fig:D3KS}(a). 
It was obtained by ensemble averaging over the weak edge-localized-mode phases achieved by the resonant magnetic perturbation field. 
Repeated transport bifurcations between the two-step staircase and the normal single-step pedestal were observed. 

In KSTAR L-mode plasmas \cite{Choi:2019wy, ChoiPPCF2024}, the two-dimensional radial corrugation of electron temperature was measured as shown in Figure~\ref{fig:D3KS}(b). 
It was observed in the phase where the MHD instabilities were suppressed and ballistic heat flux propagation events of various sizes, similar to avalanches in the near-marginal regime, are dominant \cite{Choi:2019wy, ChoiPPCF2024}.  
The dynamics of the temperature corrugation could be investigated using the high resolution image data, and the statistics of its characteristics could be analyzed \cite{ChoiPPCF2024} (see below). 
These transport barriers are intermittent but observed persistently.
Barriers can drift radially and their tread width can change in time \cite{Choi:2019wy, ChoiPPCF2024}. 
Also, they can impede the ballistic propagation of the heat flux until they are dissipated \cite{ChoiPPCF2024}, demonstrating their capability as transport barriers.

\begin{figure}
%\centering\includegraphics[width=5.0in]{TJW7.png}
\centering\includegraphics[width=6.0in]{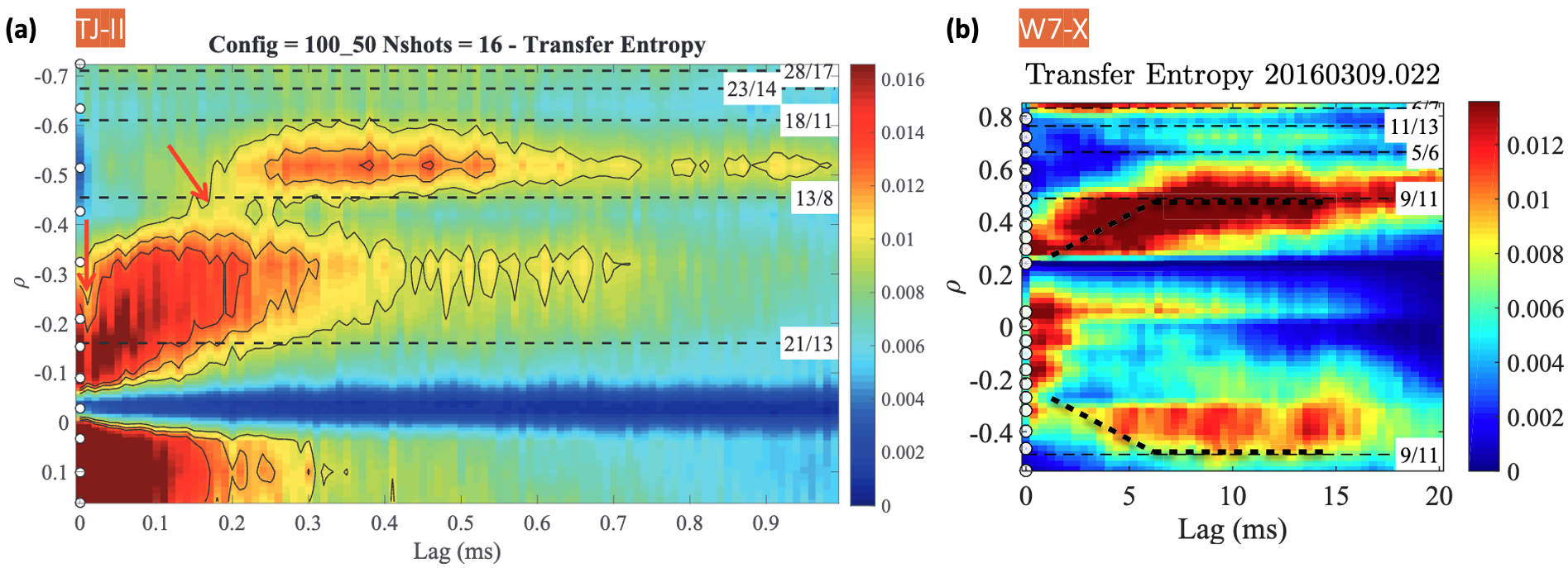}
\caption{Transfer entropy $\mathrm{TE}(\rho, \tau)$ measured in (a) the TJ-II experiment and (b) the W7-X experiment. Horizontal thin dashed lines indicate the locations of the rational surfaces of the corresponding $q$ values. Thick dashed lines indicate the propagation/trapping of the information revealed by $\mathrm{TE}(\rho, \tau)$. Red arrows indicate the immediate transfer of the information. Reprinted from Milligen et al. \cite{MilligenNF2017} (Copyright 2017 IOP) and Milligen et al. \cite{MilligenNF2018} (Copyright 2018 IOP), respectively. }
\label{fig:TJW7}
\end{figure}

The third approach is to trace the propagation of spontaneous pulses that are sufficiently mild so not to perturb transport barriers. 
In TJ-II \cite{MilligenNF2017} and W7-X \cite{MilligenNF2018} L-mode plasmas, the propagation of spontaneous and mild electron temperature perturbations was analyzed using an information theoretic measure called transfer entropy. 
The perturbations were too irregular for their propagation to be analyzed using the conventional Fourier technique.
Transfer entropy $\mathrm{TE}(\rho, \tau)$ measures the causal relation or the information transfer between two serial data separated by $\rho$ in space and $\tau$ in time, based on the degree of enhanced predictability by knowing the history of additional data \cite{SchreiberPRL2000}. 
The analysis identified the trapping (barrier) and jumping (mixing) zones of the information (see Figure~\ref{fig:TJW7}). 
Most barriers were found around the low-order rational surfaces, and they were interpreted as a result of the poloidal flow shear associated with the rational surfaces \cite{MilligenNF2023}. 
The immediate transfer of the information in the jumping zone (indicated by red arrows) was attributed to the interacting modes between different rational surfaces. 
A link between the rational surfaces and transport barriers suggests the importance of electromagnetic effects or non-adiabatic electron effects (see below). 
Barriers from different mechanisms may be not necessarily exclusive.

\subsection{Characteristics of the $\mathbf{E} \times \mathbf{B}$ staircase}
\label{sec:stairchar}

\subsubsection{The lifetime}

The $\mathbf{E} \times \mathbf{B}$ staircase can undergo merging, splitting, drifting, dissipation, and even destruction and reformation with avalanches \cite{GhendrihEPJD2014, DifNF2017, Choi:2019wy, ChoiPPCF2024}. 
In KSTAR experiments, the lifetime $t_\mathrm{life}$ of the $\mathbf{E} \times \mathbf{B}$ staircase was estimated by analyzing the high resolution spatio-temporal dynamics of the two-dimensional temperature corrugation\footnote{In other experiments, the measurement of the lifetime was technically impossible. The fast sweeping reflectometry used in Tore Supra plasmas \cite{DifPRL2015, HornungNF2017} was limited to capture the instantaneous moments of the $\mathbf{E} \times \mathbf{B}$ staircase. In HL-2A, DIII-D, TJ-II, and W7-X experiments, about 3 milliseconds (HL-2A \cite{LiuPoP2021}), 40 milliseconds (DIII-D \cite{Ashourvan:2019ek}), or 50—200 milliseconds (TJ-II \cite{MilligenNF2017} and W7-X \cite{MilligenNF2018}) records of the diagnostics data were utilized to identify the (on average) transport barriers during the analysis periods.}. 
Here, the lifetime was defined as the duration from the initial appearance through all its deformations (merging or splitting) until its disappearance.  
It appears to dissipate or be penetrated by large avalanches (see Figure 4 of reference \cite{ChoiPPCF2024}), having the lifetime of the order of 0.1—1.0 milliseconds \cite{ChoiPPCF2024}. 
Although the range of measured lifetimes was too narrow to conclude, the staircase lifetime seems to follow a long tail distribution which can be fit with a two-fold power-law function \cite{ChoiPPCF2024}. 
Note that distributions of characteristics of self-organized structures in non-equilibrium complex systems are rarely normal. 

The lower ion collisionality seems to be favorable to the robustness of transport barriers. 
A weak and scattered trend of increasing lifetime with respect to the collisionality decrease was observed within the banana regime $\nu_* < 1$ in KSTAR \cite{ChoiPPCF2024}.
In Tore Supra experiments, the $E\times B$ staircase was observed mostly in the banana regime \cite{DifNF2017}.
Note that in the GYSELA simulations the staircase is only weakly affected by the collisionality in the banana regime and it starts to be damped around the plateau transition \cite{DifNF2017}.

\subsubsection{The barrier strength}

The barrier strength was defined as the integrated zonal mean shear strength over the flow layer thickness $\delta^\mathrm{flow}$ in the GYSELA simulation \cite{DifNF2017}. 
It indicates how impermeable the flow shear layer is to the radial transport or mixing processes. 
Since the flow shear is hard to be measured accurately in experiments, a proxy of the barrier strength was measured. 
In Tore Supra experiments \cite{HornungNF2017}, the height and the width of the correlation length reduction were measured. 
Since both quantities decrease with $\rho_*^{-1}=a/\rho_s$, the barrier would be weaker as the plasma size increases. 
In HL-2A experiments \cite{LiuPoP2021}, the permeability was estimated by how the zero-time delay cross correlation between radially adjacent channels varies with the normalized fluctuation amplitude. 
The regions of enhanced and suppressed permeability (mixing) were identified. 
The enhanced permeability region coincides with the larger correlation length region.
The suppressed permeability regions are located near the correlation length minima or the pressure gradient maxima as expected. 
In KSTAR experiments \cite{ChoiPPCF2024}, the square of temperature corrugation amplitude was measured. 
This exhibited a skewed distribution, and the larger amplitude corrugation was less frequently observed. 
The squared amplitude seems to have a weak and scattered increasing trend with respect to the collisionality decrease.

\subsubsection{The tread width}
\label{sec:tread}

The tread width $\Delta$ of the $E\times B$ staircase, or the mixing zone width, corresponds to the (maximum possible) radial scale of avalanche transport. 
In Tore Supra experiments \cite{DifNF2017}, $\Delta$ was taken to be the distance between the minima of the correlation length profile.
In KSTAR experiments \cite{ChoiPPCF2024}, it was defined as the inverse of the average radial wavelength of the temperature corrugation. 
The scaling and the distribution of $\Delta$ were investigated in Tore Supra and KSTAR experiments, respectively, based on the large number of measurements.

\begin{figure}
%\centering\includegraphics[width=5.0in]{DELT.png}
\centering\includegraphics[width=6.0in]{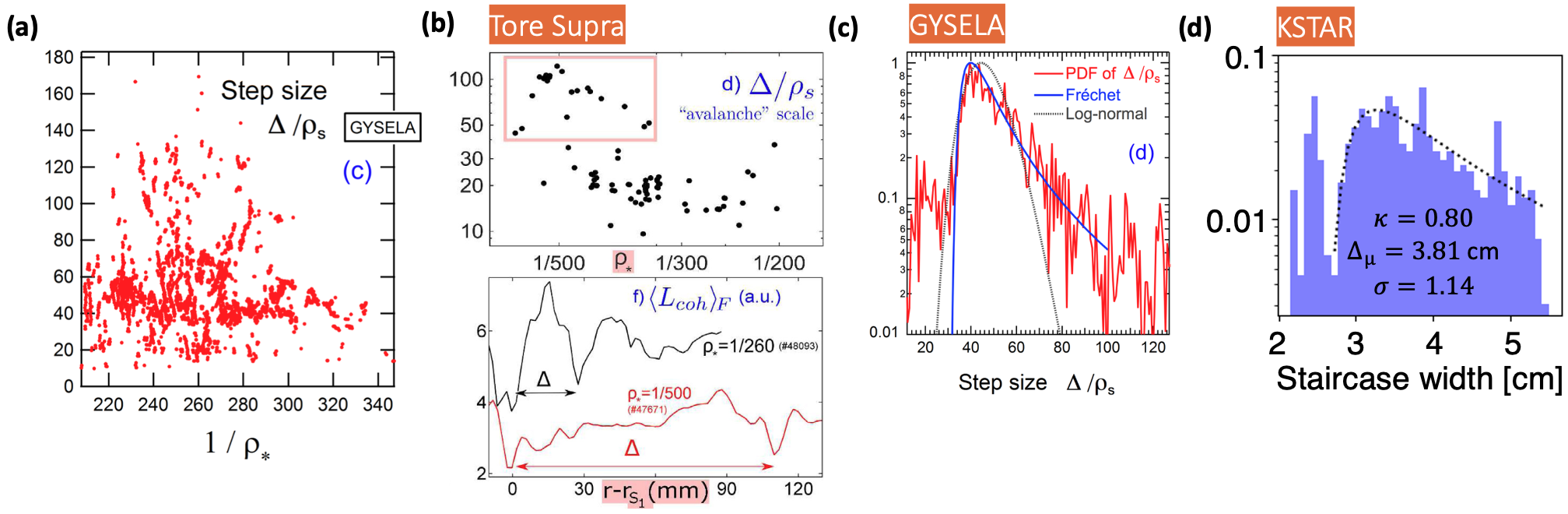}
\caption{Various measurements of the tread with $\Delta$. It is called step size in (a) and (c), avalanche (spatial) scale in (b) and staircase (tread) width in (d). (a) The normalized tread width $\Delta/\rho_s$ versus $\rho_*^{-1} = a/\rho_s$ where $\rho_s$ is the ion sound gyroradius in the GYSELA simulation. Reprinted from Dif-Pradalier et al. \cite{DifNF2017}. Copyright 2017 IOP. (b) $\Delta/\rho_s$ versus $\rho_*$, and the radial correlation length $\langle L_{coh} \rangle_F$ profile measurements for small and large $\Delta$ cases in the Tore Supra experiment. $r_{s_1}$ is the radial location of the first (expected) shear flow layer or the first $\langle L_{coh} \rangle_F$ minimum. Reprinted from Hornung et al. \cite{HornungNF2017}. Copyright 2017 IOP. The distribution of $\Delta$ in (c) the GYSELA simulation and (d) the KSTAR experiment. Here, $\kappa$, $\Delta_\mu$ and $\sigma$ are shaping, location, and scale parameters of a Fréchet distribution (dotted line) to fit the measured data in (d). Reprinted from Dif-Pradalier et al. \cite{DifNF2017} (Copyright 2017 IOP) and Choi et al. \cite{ChoiPPCF2024} (Copyright 2024 IOP), respectively.  }
\label{fig:DELT}
\end{figure}

The scaling of $\Delta$ with respect to $\rho_*=\rho_s/a$ has been of particular interest since it can be related to the global confinement scaling. 
Here, $\rho_s$ and $a$ are the ion sound gyroradius and the minor radius of plasma, respectively.
In the GYSELA simulations \cite{DifNF2017}, $\Delta$ does not seem to simply increase as $\rho_*^{-1}=a/\rho_s$ increases as shown in Figure~\ref{fig:DELT}(a). 
Instead, the most probable value has been fixed around $40~\rho_s$ for investigated values of $\rho_*$. 
This feature expects the favorable gyro-Bohm scaling. 
However, in Tore Supra experiments, $\Delta$ seems to increase with $\rho_*^{-1}$ as shown in Figure~\ref{fig:DELT}(b), implying that even with the avalanche regulation by the $E\times B$ staircase, the global confinement scaling can be like the Bohm scaling \cite{HornungNF2017}. 
Further research would be required to address this inconsistency.  

The distribution of $\Delta$ was found to follow a long tail distribution called the Fréchet distribution in the GYSELA simulations, i.e. $P(\Delta) \sim P_\text{Fréchet}(\Delta)$ \cite{DifNF2017}, and also in KSTAR experiments \cite{ChoiPPCF2024} as shown in Figures~\ref{fig:DELT}(c) and (d). 
The Fréchet distribution is one of the extreme value distributions, a distribution of maxima of samples when the samples follow a long tail distribution such as the power-law distribution. 
This Fréchet distribution can imply dual roles of avalanches, co-existing ballistic transport events (mixing process) as explained in reference \cite{ChoiPPCF2024}. 
They can destruct the $\mathbf{E} \times \mathbf{B} $ staircase but may also contribute to its formation and $\Delta$ determination\footnote{The distribution of temperature perturbation size of avalanches ($S \sim \delta T^2$ \cite{ChoiPPCF2024}) was found to follow a power-law distribution ($P(S) \sim P_\mathrm{power-law}(S)$) \cite{Choi:2019wy}. 
Assume that (1) only a temperature perturbation beyond a large size threshold can trigger the formation of the staircase and (2) $\Delta$ depends on the size of the triggering temperature perturbation which is rare and whose size would be the maximum size of perturbations for some period. Then, the Fréchet distribution of $\Delta$ follows from its definition ($P(\Delta \sim \max\{S\}) \sim P_\text{Fréchet}(\Delta)$). 
Some models have shown these behaviors, e.g., the threshold \cite{Kosuga:2013io} and the turbulence spreading strength $\sim \Delta$ \cite{AshourvanPRE2016}. 
Also, note that there are minor temperature corrugations that form away from large avalanches without apparent interactions, and their $\Delta$ distribution does not seem to follow a Fréchet distribution \cite{ChoiPPCF2024}.}.

\section{Summary and discussion}
\label{sec:sumdis}

Blobs and voids are bursty coherent structures within the ambient turbulence, which are widely observed in magnetic confinement fusion edge plasmas. 
In recent years, new progresses have been made regarding the comprehensive experimental studies of the physical behavior of blobs/voids. 
It is found that: 
\begin{enumerate}
\renewcommand{\labelenumi}{(\roman{enumi})}
\item Turbulence spreading by blobs is outward, while turbulence spreading by voids is inward, and the degree of symmetry breaking between outward propagating blobs and inward propagating voids is crucial to the total net turbulence spreading flux. 
\item The enhanced turbulence spreading induced by blobs can significantly broaden the power scrape-off width $\lambda_{q_\parallel}$ in the high collisionality regime.
\item There exist strong interactions between blobs/voids and $\mathbf{E} \times \mathbf{B}$ sheared flow. The turbulence spreading induced by blobs decreases as the flow shearing rate increases, while the turbulence spreading induced by voids is insensitive to the shearing rate. On the other hand, voids can excite drift waves and so drive intermittent zonal flow.
\end{enumerate}
These experimental findings provide novel insights and perspectives to understand the important influences of blobs and voids structures in plasma transport and core-boundary coupling issue.

% Role of avalanches in the formation and width selection of the staircase has been discussed [ChoiPPCF2024].

{\renewcommand{\arraystretch}{1.1}
\begin{table}
\caption{Experimental findings on mesoscopic transport barriers}%%%Table caption goes here
\label{table:STSM}
\resizebox{\textwidth}{!}{
\begin{tabular}{|l|l|p{3.5cm}|l|l|l|l|}%%%The number of columns has to be defined here
\hline
                      & Tore Supra \cite{DifPRL2015, HornungNF2017} & HL-2A \cite{LiuPoP2021} & DIII-D \cite{Ashourvan:2019ek} & KSTAR \cite{Choi:2019wy, ChoiPPCF2024} & TJ-II \cite{MilligenNF2017} & W7-X \cite{MilligenNF2018}   \\
\hline
Diagnostics           & Reflectometer & ECE, Reflectometer, BES & Thomson, CER & ECEI, BES & ECE & ECE \\
\hline
Measurements          & $L_c(r)$, $\mathrm{Cor}(\Delta r, \Delta t)$ & $\nabla T_\mathrm{e}(r)$, $\nabla n_\mathrm{e}(r)$, $L_c(r)$, $S(k_r, k_z)$, $\mathrm{Cor}(\Delta r, \Delta t)$ & $T_\mathrm{e}(r)$, $n_\mathrm{e}(r)$, $T_\mathrm{i}(r)$  & $\delta T_\mathrm{e}(r,z)$, $S(k_r, t)$, $\mathrm{Cor}(\Delta r, \Delta z)$ & $\mathrm{TE}(\rho, \tau)$, $\mathrm{Cor}(\Delta r, \Delta t)$ & $\mathrm{TE}(\rho, \tau)$, $\mathrm{Cor}(\Delta r, \Delta t)$ \\
\hline
Lifetime [ms]         &      &      &      & 0.1--1.0 (power-law dist.) &    &    \\
\hline
Strength              & Decrease with $\rho_*^{-1}$ &    &    & Decrease with $\nu_\mathrm{i}$ &  &  \\
\hline
Tread width $\Delta$ [$\rho_s$] & 10--100 ($\Delta  \propto \rho_*^{-1}$) & $\sim$10 & $<$10 & 10--50 (Fréchet dist.) & $\sim$10            & $\sim$50 \\ 
\hline
Rational surface      &    & $q=2, 5/2$ &  & $q=2$ & $q=3/2, 8/5, 13/8, \dots$ & $q=4/5, 9/11, \dots$\\
\hline
\end{tabular}
}
\vspace*{-4pt}
\end{table}%%%End of the table
}

The $\mathbf{E} \times \mathbf{B}$ staircase, or more generally mesoscopic transport barriers, represents a realization of inhomogeneous mixing in fusion plasmas. 
Various methods have been utilized to identify these elusive transport barriers in fusion core plasmas. 
As summarized in Table~\ref{table:STSM}, they have a lifetime order of 0.1--1.0~ms with a long tail distribution (KSTAR), and a tread width of 10--100~$\rho_s$ (Tore Supra, HL-2A, TJ-II, W7-X) with a Fréchet distribution (KSTAR). 
Their strength seems to decrease with $\rho_*^{-1}$ (Tore Supra) or $\nu_\mathrm{i}$ (KSTAR), and they seem to be associated with the low-order rational surfaces (HL-2A, KSTAR, TJ-II, W7-X). 

It may be worth discussing potentially complex effects of the low-order rational surfaces on plasma inhomogeneity or the $\mathbf{E} \times \mathbf{B}$ staircase. 
On the one hand, the low-order rational surface can be a region of enhanced mixing \cite{Waltz:2006hs, RathPoP2021}.
The transport flux by more densely packed $m/n$ modes at the low-order rational surface can be relatively higher than that in adjacent regions, leading to the local flattening of profiles \cite{Waltz:2006hs}. 
The low-order rational surface can be an initiation location of avalanche-like transport events \cite{Choi:2019wy, RathPoP2021}.
On the other hand, the generation of zonal flows out of turbulence self-interactions can be also more facilitated at the low-order rational surface \cite{DominskiPoP2015} since eddies can bite their own tail more easily with the shorter parallel length \cite{brioschi2025}.  
The existence of rational surfaces can signify the ubiquitous inhomogeneous mixing (corrugated profiles) in magnetically confined toroidal plasmas.

According to the recent simulation \cite{brioschi2025} and theory \cite{HahmPoP2023, ChoiNF2024}, the inhomogeneous self-generation of zonal flow at the rational surface can be more effective with the higher fraction of the fast ions, lowering the zonal flow generation threshold \cite{HahmPoP2023, ChoiNF2024}. 
This implies the stronger plasma inhomogeneity in future fusion experiments with significant energetic ions.  
Interestingly, avalanches and mesoscopic transport barriers in the electron channel were readily observed in the low-level-MHD phase of the KSTAR fast-ion regulated enhancement (FIRE) mode \cite{Lee2025FIRE}, characterized by the high fast ion fraction \cite{HanNat2022}
The barriers seem to have a longer lifetime in the FIRE mode compared to the similar L-mode. 
If long stationary mesoscopic transport barriers could be realized in future experiments, it would be an attractive confinement regime which is free from disruptive MHD instabilities.

Although some progress has been made over the past decade in the experimental side, a comprehensive understanding of the physics of the $\mathbf{E} \times \mathbf{B}$ staircase remains elusive.
A brief introduction of existing specific models on the $\mathbf{E} \times \mathbf{B}$ staircase is provided in Appendix.

In the physics literature, there are three classes of models which address staircase formation.
These are (i) flux bistability, (ii) homogenization, and (iii) phase separation (spinodal decomposition).
These are explained below.

\begin{figure}
%\centering\includegraphics[width=5.0in]{PD03.png}
\centering\includegraphics[width=6.0in]{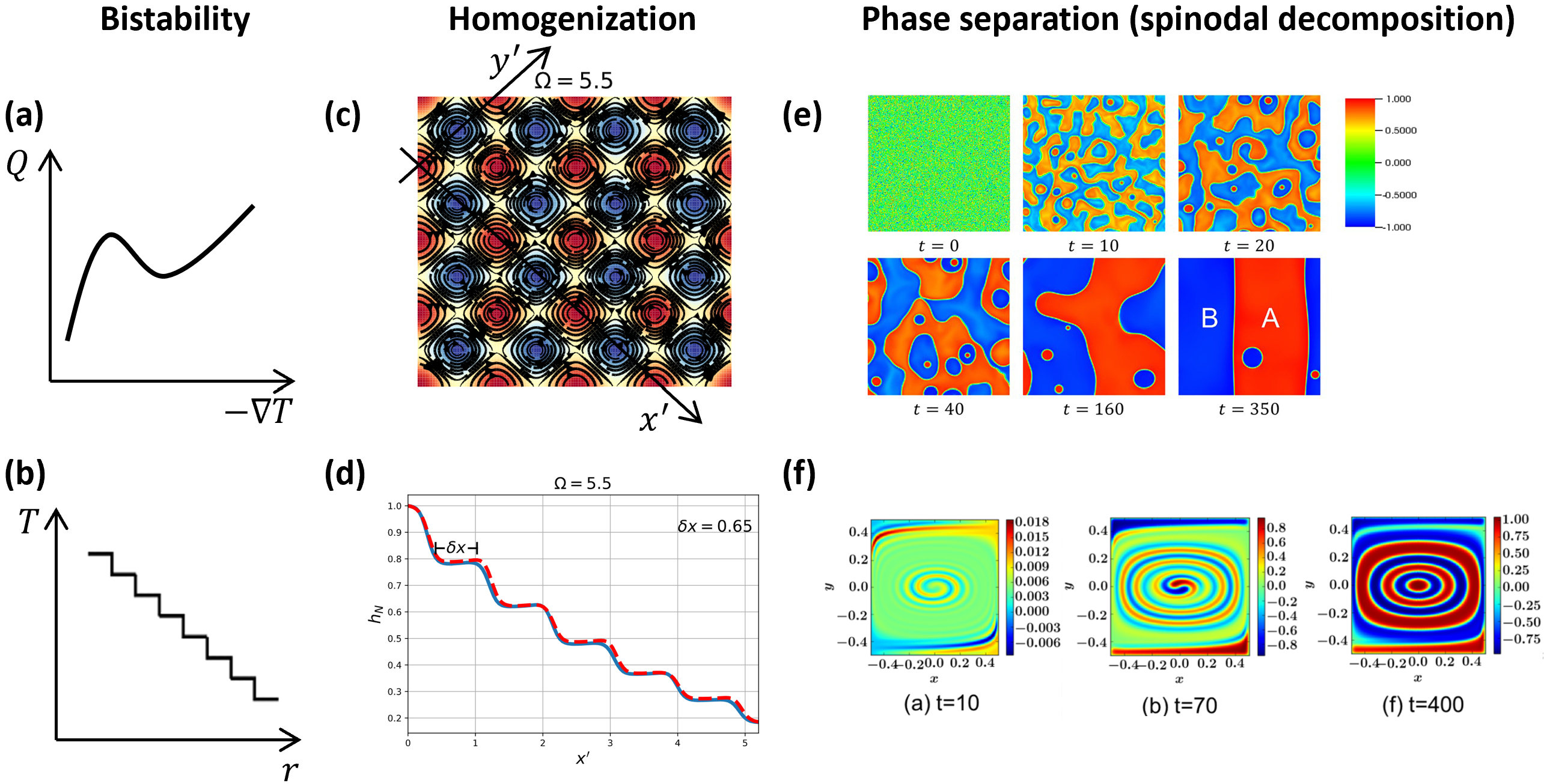}
\caption{(a) S-curve relating heat flux $Q$ and temperature gradient $\nabla T$. (b) Staircase profile of temperature $T$. (c) An array of vortex cells and (d) the averaged (in $y^\prime$ direction) scalar concentration profile $h(x^\prime)$ in the form of the staircase. Reprinted from Ramirez and Diamond \cite{RamirezPRE2024}. Copyright 2024 APS. (e) An illustration of phase separation in the binary fluid. Colors indicate the normalized component density contrast $\psi = (\rho_{A} - \rho_{B})/(\rho_{A} + \rho_{B})$. Reprinted from Fan et al. \cite{FanPoP2018}. Copyright 2018 AIP. (f) Evolution of $\psi$ patterns from initial spirals to concentric annuli (staircase). Reprinted from Fan et al. \cite{FanPRE2017}. Copyright 2017 APS.}
\label{fig:PD03}
\end{figure}

\begin{enumerate}
\renewcommand{\labelenumi}{(\roman{enumi})}
\item Flux bistability depends on the coexistence of two different values of the local gradient at the same value of the total flux. Of course, the total flux is set by the sum of turbulent and collisional (neoclassical) fluxes, and the ratio of these two varies considerably. Bistability is described by the familiar S-curve for the highly nonlinear flux-gradient relation (see Figure~\ref{fig:PD03}(a)). This is familiar from studies of single barriers. However, a staircase is an array or lattice of (small) barriers, with regions of strong mixing interspersed between (see Figure~\ref{fig:PD03}(a)) \cite{DritschelJAS2008, BalmforthJFM1998}. The classic example of a bistable mixing model of staircases is that of Balmforth, Llewellyn-Smith and Young (hereafter BLY) \cite{BalmforthJFM1998}, which has yielded a bounty of insights into staircase structure and evolution. The BLY model aimed to explain the observations of density staircases in stable-but-buoyant mixing. The key feature of BLY is its use of two mixing lengths in an otherwise standard K-model approach. In BLY, the mixing scale $l$ is related to an imposed stirring scale $l_{ext}$ and an emergent scale $l_{oz}$, the Ozmidov scale \cite{OzmidovIAOP1965}, by $1/l = 1/l_{ext} + 1/l_{oz}$. The Ozmidov scale is dependent upon turbulence intensity and the density gradient, and is defined by the cross-over of the buoyancy frequency and the eddy turn over rate. The flux-gradient relation for the system exhibits a structure similar to the S-curve, discussed above. The density profile consists of an array of strong mixing zones interleaved by steep gradient layers (barriers).

\item Homogenization refers to the well-known tendency of eddies with differentially rotating flow to homogenize the profile of vorticity, scalar concentration, etc. within the eddy. Then, an array of tangential eddies, as shown in Figure~\ref{fig:PD03}(c), in a fluid with molecular diffusivity $D_0$ and large Peclet number $Pe = vl/D_0 \gg 1$, naturally leads to inhomogeneous mixing \cite{WeissPRS1966,RamirezPRE2024}. This is because mixing within the eddy is much faster than the slow diffusion through the separatrices between eddies. Thus, scalar concentration profiles naturally take the form of staircases, i.e. flat within the eddies and steep in the regions between them, as in Figure~\ref{fig:PD03}(d).

\item Phase separation \cite{PadhanJFM2025} refers to the property of some mixtures which allows them to separate into domains of different constituent components (see Figure~\ref{fig:PD03}(e) \cite{FanPoP2018}). Such a process also goes by the name of spinodal decomposition. Phase separation can lead to layering, as in Figure~\ref{fig:PD03}(f) \cite{FanPRE2017}, where a target pattern results, formed by alternating bonds of different phases.
\end{enumerate}

The results of recent blobs/voids research can provide helpful insights into developing a better physical model for the $\mathbf{E} \times \mathbf{B}$ staircase (avalanches and zonal flow layers).
Incorporating turbulence intensity flux and its dynamics into physical models of the $\mathbf{E} \times \mathbf{B}$ staircase is necessary.
In fact, such models have been developed but have not yet been exercised effectively. 
In particular, the length scale for the turbulence intensity can differ from that for the transport flux, adding a scale to the bistable mixing model \cite{AshourvanPRE2016}.
This is plausible, since research on coherent structures \cite{long2024_structures} indicates a departure of the spreading flux from the particle flux. 
The scale of the spreading flux effectively sets the scale of the turbulence envelope, which in turn determines the zonal flow.

Figure~\ref{fig:TL02} shows that the symmetry breaking of turbulence intensity flux, carried by the coherent structures such as blobs/voids, yields a net turbulence spreading and forms a region of enhanced mixing. 
On the other hand, recent studies \cite{sladkomedova2023,dumeratPPCF2025} introduced in item~\ref{sec:blobflow} of Section~\ref{sec:blob} show that the blobs/voids can contribute to the zonal flow generation in core plasmas as they propagate across the plasma boundary\cite{DiamondPPCF2008, HahmPoP2024, CaoPRL2025}. 
The resulting sheared flow layer forms a region of suppressed mixing. 
The correlated growth of zonal flows at distant regions might be mediated by turbulence spreading of momentum flux between zonal flow layers \cite{ZhuPRL2020, MilovanovPRE2021}. 
Therefore, relating the coherent structures and their propagation to the formation or sustainment of the $\mathbf{E} \times \mathbf{B}$ staircase should not be surprising. 
Indeed, some relevant observations were already reported, i.e. the triggering of temperature staircase by the excessive localized heating \cite{ChoiPPCF2024}.

A key question for the $\mathbf{E} \times \mathbf{B}$ staircase is the scale selection problem -- i.e. what sets the separation between zonal flow layers? 
There are several possibilities:

\begin{enumerate}
\renewcommand{\labelenumi}{(\roman{enumi})}
\item Bistable mixing (two scales model). Here, the emergent scale adjusts to define the layer separation. This is an outcome of the nonlinear flux-gradient relation. In the case of drift-Rossby turbulence, the Rhines scale \cite{RhinesJFM1975} is selected (see Appendix).

\item Blob/void emission. In this scenario, a propagating structure (i.e. blob or void) emits (Cherenkov) drift waves, much like a test particle. The drift-waves determine a Reynolds stress which in turn drives a zonal flow, and thus a shear layer. This shear layer regulates the structure. The length from birth-of-structure to shearing-of-structure defines a scale, which can set the separation between staircase layers (i.e. the tread width). This may be thought of as a `mean free path to shearing' length. Such a scale depends fundamentally on structure induced mixing.  

\item Turbulence spreading. More generally, the length scale for turbulence spreading will set a turbulence envelope scale, which determines the Reynolds force scale and thus the shear layer spacing.  
\end{enumerate}

Further research -- experimental, theoretical and computational -- is necessary to sort out which mechanism is relevant. 
Experimental research includes, but is not limited to, (1) the multi-machine analysis, (2) the development of a stable staircase scenario (possibly with the help of energetic particles \cite{HahmPoP2023, ChoiNF2024} or light impurities \cite{SeoPoP2022}), and (3) the development of fine-scale flow or potential diagnostics, which enables the detailed analysis of inhomogeneous mixing process (see Section~\ref{sec:blob}), in the core plasma.

\section*{Appendix}

\subsection{Production ratio for gyrokinetic ITG model}

The basic equation is the gyrokinetic equation and the relevant intensity is $\langle g^2 \rangle$, where $g$ is the non-adiabatic distribution function. 
We briefly sketch a derivation for the simplified case of electrostatic ion drift kinetics in a straight field.
The basic equation is:
\begin{eqnarray}
\frac{\partial g}{\partial t} + \mathbf{v}_\perp \cdot \nabla g = \frac{|e|}{T_\mathrm{i}} \frac{\partial \phi}{\partial t}  \langle f \rangle + v_{*\mathrm{i}}(E) \frac{\partial }{\partial y} \left( \frac{|e|\phi}{T_\mathrm{i}} \right) \langle f \rangle 
\end{eqnarray}
Here, $\langle f \rangle$ is the inhomogeneous mean distribution and $v_{*\mathrm{i}}(E)$ is the energy-dependent full diamagnetic velocity (including temperature gradient dependent piece).

Multiplying by g gives:
\begin{eqnarray}
\frac{\partial \langle g^2 \rangle}{\partial t} + \nabla \cdot \langle \mathbf{v}_\perp g g \rangle = \frac{|e|}{T_\mathrm{i}} \langle f \rangle \left[ \left\langle \frac{\partial \phi}{\partial t} g \right\rangle + v_{*\mathrm{i}}(E) \left\langle \frac{\partial \phi}{\partial y} g \right\rangle \right]
\end{eqnarray}
The kinematic production ratio then is the comparison of the velocity integrated second term to that for the third term.

Integrating over velocity space gives (for a steady state):
\begin{eqnarray}
\nabla \cdot \int d^3v \langle \mathbf{v}_\perp g g \rangle = \int d^3v \frac{|e|}{T_\mathrm{i}} \langle f \rangle \left[ \left\langle \frac{\partial \phi}{\partial t} g \right\rangle + v_{*\mathrm{i}}(E) \left\langle \frac{\partial \phi}{\partial y} g \right\rangle \right]
\end{eqnarray}
The production ratio can be given by 
\begin{eqnarray}
PR = d\left(\int d^3v \langle \tilde{v}_\perp g g \rangle \right) / \mathcal{R}
\end{eqnarray} 
where $\mathcal{R} = \int_{r-\delta}^{r+\delta} dr \int d^3v \frac{|e|}{T_\mathrm{i}} \langle f \rangle \left[ \left\langle \frac{\partial \phi}{\partial t} g \right\rangle + v_{*\mathrm{i}}(E) \left\langle \frac{\partial \phi}{\partial y} g \right\rangle \right]$ and $d(F) \equiv F(r+\delta) - F(r-\delta)$.
Note that $PR$ is still the ratio of the differential of the spreading flux in phase space to the integrated production due to phase space relaxation. 
In practice, the kinetic $PR$ here can be calculated only via simulation.

\subsection{Models on the $\mathbf{E} \times \mathbf{B}$ staircase}

Many theoretical models have been suggested to understand the physics of the $\mathbf{E} \times \mathbf{B}$ staircase. 
The $\mathbf{E} \times \mathbf{B}$ staircase was characterized as a mesoscopic transport state with co-existing super-diffusive transport events (avalanches) and permeable transport barriers (zonal flow layers) \cite{DifNF2017}. 
It was found in different frameworks \cite{GarbetPoP2021, MilovanovPRE2021, ChenNF2025} that such a state can be realized, though their physical interpretations may differ. 
In Garbet et al. \cite{GarbetPoP2021}, avalanches and zonal flow layers are the radially propagating wave packets and the wave trapping \cite{SmolyakovPRL2000, SasakiPoP2018, SasakiPoP2021} quasi-periodic zonal flow solution in the system, respectively. 
In Milovanov et al. \cite{MilovanovPRE2021}, zonal flow layers were described by coupled oscillators of nonlinear Schrödinger equation whose vibrations and sub-quadratic interaction allow a super-diffusive radial transport. 
In Chen et al. \cite{ChenNF2025}, drift wave solitons (akin to avalanches) can develop via the nonlinear drift wave and zonal flow interaction, and they are confined by spontaneously excited zonal flows. 

Some researches \cite{Kosuga:2013io, QiNF2022, AshourvanPRE2016, GuoPPCF2019, WangNF2018, LecontePoP2021, YanNF2022} have emphasized a role of mixing processes in the formation of mesoscopic zonal flows or profile corrugation. 
In Kosuga et al. \cite{Kosuga:2013io}, the pressure corrugation was understood as the jamming of avalanching heat flux due to the finite time delay \cite{QiNF2022}. 
In Ashourvan and Diamond \cite{AshourvanPRE2016} and Guo et al. \cite{GuoPPCF2019}, the bistability by dynamic mixing length was adopted to implement the inhomogeneous mixing and result in the density corrugation. 
In Wang et al. \cite{WangNF2018}, the initial zonal flow driven by the Reynolds stress was fortified by the in-phase mean radial electric field variation during the temperature relaxation event. 
In Leconte and Kobayashi \cite{LecontePoP2021}, the radial variation of the transport cross phase between the density and electric potential fluctuations (and the associated particle flux) can result in the density corrugation.  
In Yan and Diamond \cite{YanNF2022}, alternating the enhanced and mitigated transport regions via the wave-particle resonance and non-resonance, respectively, due to the spatial structure of the mean flow lead to a corrugated profile in the trapped ion mode turbulence. 

It is also noteworthy that a role of rational surface with kinetic electrons on the zonal flow was investigated in other researches (Figure~\ref{fig:TJW7} and also experiments in HL-2A \cite{LiuPoP2021} and KSTAR \cite{Choi:2019wy} implied a role of the rational surface). 
In Dominski et al. \cite{DominskiPoP2015}, the passing electron destabilization of turbulence and the following zonal flow at the rational surface were revealed. 
In Rath et al. \cite{RathPoP2021}, the lock-in of zonal mode at the rational surface was observed.

\subsection{Some lessons from analogies}

Revisiting details of specific models and comparing them with experimental results are left for future work. 
Here, in line with the theme of the workshop, “Layering—A structure formation mechanism in oceans, atmospheres, active fluids and plasmas”, the $\mathbf{E} \times \mathbf{B}$ staircase is viewed as one of self-organized layering phenomena in non-equilibrium complex systems and compared with other long-studied phenomena. 
Although there may be no universal laws for self-organized structures in complex systems, lessons can be learned by comparing observations in different systems \cite{GoldenfeldScience1999}. 

One well known example of mesoscopic structures in non-equilibrium system is the Rayleigh-Bénard (RB) convection cells \cite{Livi2017}. 
It occurs in a fluid between two horizontal plates at different temperatures when the bottom plate is sufficiently hotter than the top plate ($\Delta T = T_\mathrm{bottom} - T_\mathrm{top} > \Delta T_\mathrm{crit}$).
The physical meaning of this condition lies in the competition between conductive and convective transport.  
The ratio between the conductive and convective time scales is the Rayleigh number which depends on $\Delta T$, i.e. $\frac{\tau_\mathrm{cond}}{\tau_\mathrm{conv}} = Ra(\Delta T)$. 
The convection cells appear when the convective transport is sufficiently more efficient than the conductive transport ($\tau_\mathrm{conv} \ll \tau_\mathrm{cond}$), overcoming the viscous force.
It is the transition from a homogeneous disordered conductive state to an inhomogeneous ordered convective state.

The $\mathbf{E} \times \mathbf{B}$ staircase is a pattern in the near-marginal regime characterized by the Kubo number ($Ku = \tau_\mathrm{corr} / \tau_\mathrm{turn}$, the ratio between the correlation time and the eddy turnover time) larger than 1.
$Ku > 1$ means more efficient eddy-particle interaction than decorrelation or detrapping processes, and results in the discrepancy from the local diffusive transport models \cite{GillotPPCF2023}. 
Both the $\mathbf{E} \times \mathbf{B}$ staircase and the RB convection cells are patterns that form away from a state dictated by the homogeneous mixing of diffusive process (see Table~\ref{table:STAN}).
It demands us to go beyond the diffusive paradigm \cite{HahmJKPS2018}. 

However, the $\mathbf{E} \times \mathbf{B}$ staircase is the secondary structure out of turbulent fluctuations, while the RB cells are more like steady laminar flows in the pattern of regular cells driven by the primary instability. 
While the scale of the RB convection cells is largely determined by the most unstable linear mode in the system, the nonlinear dynamics would matter for the scales of the $\mathbf{E} \times \mathbf{B}$ staircase, $\delta^\mathrm{flow}$ and $\Delta$. 
For example, $\delta^\mathrm{flow}$ can be extended by the turbulence spreading \cite{YiPoP2024}.

As the $\mathbf{E} \times \mathbf{B}$ staircase can be destroyed by avalanches (turbulent heat flux), the RB cells also become more complex, deformed or destructed, with turbulent fluctuations as the temperature difference is further increased (or, equivalently the viscosity is reduced).
In Ramirez and Diamond \cite{RamirezPoP2025}, the effect of turbulent fluctuation on a staircase state of cellular flows was investigated, providing insight into the fluctuation strength condition that the staircase state can survive.

{\renewcommand{\arraystretch}{1.1}
\begin{table}
\small 
\caption{Analogies}%%%Table caption goes here
\label{table:STAN}
\resizebox{\textwidth}{!}{
\begin{tabular}{| p{2cm} | p{4cm} | p{4cm} | p{4cm} |}%%%The number of columns has to be defined here
\hline
 & Rayleigh-Bénard cells & Zonal jets & The $\mathbf{E} \times \mathbf{B}$ staircase \\
\hline
Description & Convective cells & Inhomogeneous potential vorticity (PV) mixing & Zonal flow layers intervening in avalanches \\
\hline
Key condition & $Ra = \frac{\tau_\mathrm{cond}}{\tau_\mathrm{conv}} > Ra_\mathrm{crit}$ & $\tau_\mathrm{wave} \sim \tau_\mathrm{turn}$\footnotemark[4] & $Ku = \frac{\tau_\mathrm{corr}}{\tau_\mathrm{turn}} > 1$ \footnotemark[5]\\
\hline
Meaning of key condition & Efficient convection overcoming viscosity & Balance between linear waves dynamics and nonlinear eddies dynamics & Efficient eddy-particle interaction overcoming decorrelation or detrapping processes \\
\hline
Positive feedback & Rayleigh-Bénard instability & PV gradient $\sim$ Rossby elasticity & ? \\
\hline
Scale selection & $\sim$Most unstable linear mode & The Rhines scale & ? \\
\hline
\end{tabular}
}
\vspace*{-4pt}
\end{table}%%%End of the table
}
\footnotetext[4]{Equating an eddy turnover time $\tau_\mathrm{turn}$ to the Rossby wave period $\tau_\mathrm{wave}$ yields a transition scale. Depending on how we define a turnover time, different scales can be found \cite{VallisJPO1993}. The Rhines scale \cite{RhinesJFM1975} is obtained when $\tau_\mathrm{turn}$ is given by $\lambda_\mathrm{corr} / v$ where $\lambda_\mathrm{corr}$ and $v$ are the correlation length and typical velocity of the fluid eddy.}
\footnotetext[5]{Within a strict definition based on the original reference \cite{DifPRE2010}, the $\mathbf{E} \times \mathbf{B}$ staircase is not merely a set of simple shear flow layers, but rather shear flow layers that co-exist with avalanche transport. In other words, they are shear flow layers that originate and evolve in close relationships with avalanche transport. $Ku>1$ condition, indicative of the avalanche transport regime, would be a necessary but not sufficient condition for the $\mathbf{E} \times \mathbf{B}$ staircase. Note that this does not mean that zonal flows cannot exist in $Ku<1$ regime \cite{MarstonPRL2016, MarstonARFM2023}.}

A more relevant and famous analogy has been made with zonal jets in the geophysical fluid \cite{DiamondPPCF2005, GurcanJPA2015}. 
Indeed, the formal analogy can be made between the Charney equation for the (beta-plane) geophysical fluid dynamics and the Hasegawa-Mima equation for the (slab geometry) drift wave turbulence in plasmas. 

In those simplified systems, there exists a conserved quantity called potential vorticity (PV; $Q$) attached to a fluid element, i.e. $\frac{dQ}{dt} \approx 0$ where $d/dt$ indicates the material derivative. 
For the total PV variation is preserved at zero within the system, if the PV mixing (by turbulent eddies or wave breaking) occurred in one location and reduce the PV gradient there, the PV gradient should be increased in other place. 
This is called the inhomogeneous PV mixing or PV staircase \cite{DritschelJAS2008}.

According to the Rhines’ argument, the width of the PV mixing region can be limited by the Rhines scale $L_\mathrm{Rh}$ \cite{RhinesJFM1975} representing the balance between the linear Rossby wave dynamics and the nonlinear eddy dynamics. 
Below $L_\mathrm{Rh}$, the dominant nonlinear eddy dynamics can lead to a fully developed 2D turbulence, mixing PV and resulting in the inverse cascade of energy toward $L_\mathrm{Rh}$. 
The initial mixing can be facilitated by the reduction of the PV gradient ($\sim$the Rossby elasticity) that favors instabilities and eddies rather than waves. 
The inverse cascade stops near $L_\mathrm{Rh}$ above which the linear Rossby wave dynamics dominates. 
The steep PV gradient can form between two PV mixed regions, supporting and guiding Rossby waves. 
Zonal jets can emerge via the triad interactions in Rossby wave turbulence including the zonal mode \cite{HagimoriPoF2024}, sharpening the PV gradient further and resulting in the PV staircase. 

One enlightenment from the analogy with the zonal jets is the non-acceleration theorem \cite{CharneyJGR1961}.
It states that steady and ideal (non-dissipative) waves or fluctuations have no net effect in changing the mean flow of the system, implying the important role of turbulence spreading flux (see item \ref{sec:blobspread} of Section~\ref{sec:blob}) in accelerating the zonal flow \cite{DiamondPPCF2008, HahmPoP2024}. 
For example, the turbulence influx through the plasma boundary can play a role to drive zonal flows in the core plasma (see item \ref{sec:blobflow} of Section~\ref{sec:blob}).
Turbulence spreading can mediate the interaction between zonal flow layers \cite{ZhuPRL2020} of the $\mathbf{E} \times \mathbf{B}$ staircase, suggesting an explanation for its long range order.

In realistic fusion plasmas, however, directly adopting the conventional picture of the inverse cascade is debatable. 
It relies on the large separation between the injection scale and the dissipation scale, but these scales may not be clearly distinguished in fusion plasmas with multi-scale instabilities. 
Also, there can be multiple processes generating a radial mixing zone in fusion plasmas \cite{GarbetNF1994, HahmPPCF2004, KishimotoPTRA2023}, and defining the corresponding scale to the Rhines scale is not trivial. Avalanches are intrinsically the scale-free transport \cite{DiamondPoP1995} unless they are limited by zonal flows. 
The width of the radial mixing region $\Delta$, the radial transport scale or the distance between the zonal flow layers, might be rather set by the competition or interaction between the radial (avalanches) and poloidal (zonal flows) structure formation processes \cite{DiamondNF2001}. 
Since zonal flows can be facilitated or locked in at the rational surface \cite{DominskiPoP2015, RathPoP2021, brioschi2025}, it might be also necessary to take the plasma current profile and its external constraints into account.

\section*{Acknowledgments}

The authors would like to thank the Isaac Newton Institute for Mathematical Sciences for the support and hospitality during the programme (Anti-diffusive dynamics: from sub-cellular to astrophysical scales (ADI)) when work on this paper was discussed. 
The authors would like to thank Rui Ke, Mingyun Cao, A. Sladkomedova, Ting Wu, Xu Chu, Rameswar Singh, Jinbang Yuan, T.S. Hahm, and G. Dif-Pradalier for many enlightening discussions.

This work was supported by: 
the Ministry of Science and Technology of the People’s Republic of China under Grant No. 2024YFE03190004, 2022YFE03100004; 
the U.S. Department of Energy, Office of Science, Office of Fusion Energy Sciences under Award No. DE-FG02-04ER54738; 
the Sci DAC ABOUND Project, scw1832; 
the EPSRC under Grant No. EP/R014604/1; 
the National Natural Science Foundation of China under Grant Nos. 12575237 and 12305238; 
the Nuclear Technology R\&D Program under Grant No. HJSYF2024(02); 
the Science and Technology Department of Sichuan Province under No. 2025ZNSFSC0059; 
the Southwestern Institute of Physics Project under No. 202301XWCX001-02; 
R\&D Programs of ``High Performance Tokamak Plasma Research \& Development (code No. EN2601-17)'', ``Korea-US Collaboration Research for High Performance Plasma on Tungsten Divertor (EN2603-02)'', and ``High Performance Fusion Simulation R\&D (EN2641-12)'' through the Korea Institute of Fusion Energy (KFE) funded by the Government funds, Republic of Korea;
the National Research Foundation (NRF) of Korea under Grant No. RS-2023-00281272.

\section*{Author contributions}

Ting Long: conceptualization, formal analysis, funding acquisition, investigation, methodology, project administration, validation, writing—original draft, writing—review and editing, visualization, data curation, resources, software.
Minjun J. Choi: conceptualization, formal analysis, funding acquisition, investigation, methodology, project administration, supervision, validation, writing—original draft, writing—review and editing, visualization, data curation, resources, software.
Patrick H. Diamond: conceptualization, formal analysis, funding acquisition, investigation, methodology, project administration, supervision, validation, writing—original draft, writing—review and editing.

All authors gave final approval for publication and agreed to be held accountable for the work performed therein.

%%%%%%%%%% Insert bibliography here %%%%%%%%%%%%%%

\bibliographystyle{RS}
% \bibliography{staircase}

\end{document}